\documentclass[conference, onecolumn]{IEEEtran}
\usepackage{graphics} 
\usepackage{epsfig} 
\usepackage{mathptmx} 
\usepackage{times} 
\usepackage{amsmath} 
\usepackage{amssymb}  
\usepackage{url}
\usepackage{color}
\usepackage{subfig}
\newtheorem{theorem}{Theorem}
\newtheorem{corollary}{Corollary}

\newtheorem{example}{Example}

\begin{document}
\title{Harnessing Bursty Interference in Multicarrier Systems with Feedback}
\author{
  \IEEEauthorblockN{Shaunak Mishra\\}
  \IEEEauthorblockA{UCLA \\ Email: shaunakmishra@ucla.edu\\} 
  \and
  \IEEEauthorblockN{I-Hsiang Wang \\}
  \IEEEauthorblockA{NTU \\
    Email: ihwang@ntu.edu.tw \\}
  \and
  \IEEEauthorblockN{Suhas Diggavi\\}
  \IEEEauthorblockA{UCLA \\
    Email: suhasdiggavi@ucla.edu}
}
\maketitle
\begin{abstract}
We study parallel symmetric $2$-user interference channels when the interference is bursty and feedback is available from the respective receivers. Presence of interference in each subcarrier is modeled as a memoryless Bernoulli random state. The states across subcarriers are drawn from an arbitrary joint distribution with the same marginal probability for each subcarrier and instantiated i.i.d. over time. For the linear deterministic setup, we give a complete characterization of the capacity region. For the setup with Gaussian noise, we give outer bounds and a tight generalized degrees of freedom characterization. We propose a novel helping mechanism which enables subcarriers in very strong interference regime to help in recovering interfered signals for subcarriers in strong and weak interference regimes. Depending on the interference and burstiness regime, the inner bounds either employ the proposed helping mechanism to code across subcarriers or treat the subcarriers separately.
The outer bounds demonstrate a connection to a subset entropy inequality by Madiman and Tetali \cite{madiman_tetali}.
\end{abstract}
\IEEEpeerreviewmaketitle

\section{Introduction} \label{sec:introduction}
The temporal nature of interference in wireless networks depends on the underlying traffic as well as the subcarrier allocations of neighbouring base stations (which usually employ multicarrier systems like OFDM). In practice, due to the bursty nature of data traffic and uncoordinated subcarrier allocations across base stations, the resulting interference at the physical layer tends to be bursty. In addition to the potential for harnessing such burstiness, feedback from the receivers is another resource available in wireless networks. With these motivations, in this paper we study parallel (multicarrier) interference channels with bursty interference links and output feedback from the receivers.

\par In \cite{KPV_ISIT09} and \cite{KPV_ITW09}, the problem of harnessing bursty interference was studied for a single carrier setup without feedback. A multicarrier version of \cite{KPV_ISIT09} was studied in \cite{self_isit13_w}. To study benefits of feedback, \cite{ihsiang_bursty_feedback} considered a single carrier setup with bursty interference and output feedback from the receivers. In \cite{ihsiang_bursty_feedback}, bursty interference was modeled using a Bernoulli random state (instantiated i.i.d. over time) and a complete capacity characterization was given for the linear deterministic setup. In this paper, we study the multicarrier version of \cite{ihsiang_bursty_feedback} \emph{i.e.,} output feedback in multicarrier systems with bursty interference. Since \cite{ihsiang_bursty_feedback} developed optimal single carrier schemes, a natural question arises in the multicarrier version: is it always optimal to treat each subcarrier \textit{separately} and just copy the optimal scheme in \cite{ihsiang_bursty_feedback} on each subcarrier? As the following example illustrates, such a separation may not be always optimal.

\begin{figure}[!ht]
\begin{center}
\includegraphics[scale=0.65]{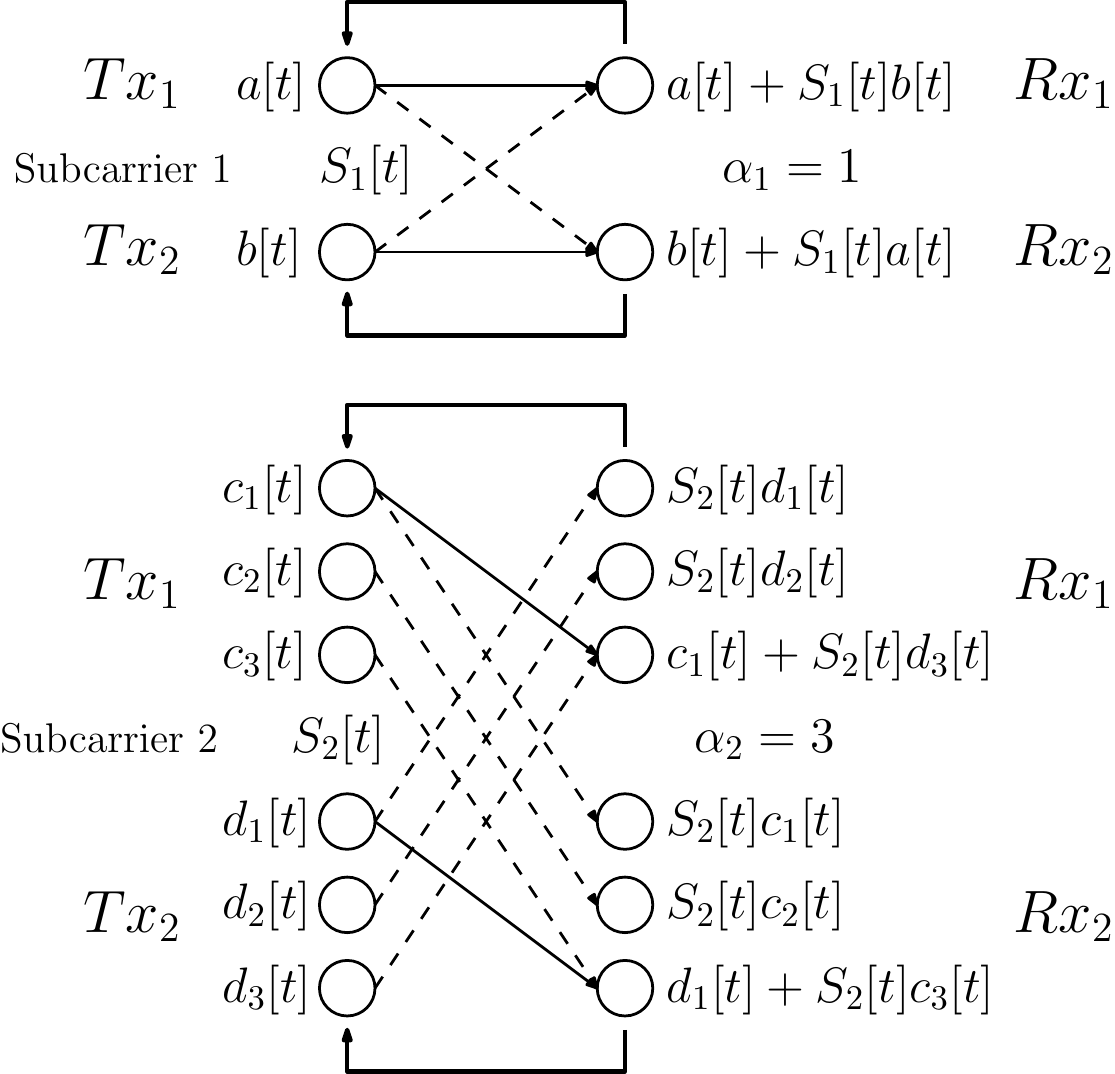}
\caption{Toy example with bursty interference in $2$ subcarriers.}
\label{fig:toy_example}
\end{center}
\end{figure}

\paragraph*{Toy example} Consider two parallel symmetric $2$-user linear deterministic interference channels (LDICs) \cite{ADT_det_channel} as shown in Figure~\ref{fig:toy_example}. The first subcarrier has one direct link ($n_1 =1$) and one interfering link ($k_1=1$, hence $\alpha_1= \frac{k_1}{n_1}=1$) and the second subcarrier has one direct link and three interfering links ($\alpha_2=\frac{k_2}{n_2}=3$). Causal output feedback is available from the receivers to the respective transmitters. Bernoulli random states $S_1[t]$ and $S_2[t]$ indicate the presence of interference in the first and second subcarrier respectively and are instantiated i.i.d. (over time) from an arbitrary joint distribution $\mathbb{P}_{S_1 S_2}$. For this example, we assume the expectation of both the states to be $p=\frac{1}{2}$.
Our goal here is to find the maximum achievable symmetric rate. Using the optimal single carrier schemes in \cite{ihsiang_bursty_feedback}, we can achieve symmetric rate $0.667$ from the first subcarrier and symmetric rate $1.25$ from the second subcarrier. Summing these rates, we can achieve a total symmetric rate $ 1.917$. Now, we will show that rate $2.0$ is achievable by coding \textit{across} the subcarriers rather than treating the subcarriers separately. We use a block based pipelined scheme (block length $N_B$) as follows. The transmitters always send fresh symbols in the first subcarrier ($a$-symbols for $Rx_1$ and $b$-symbols for $Rx_2$ as shown in Figure~\ref{fig:toy_example}). In the first subcarrier, for sufficiently large $N_B$, with high probability (w.h.p.) only $pN_B$ $a$-symbols in a block get interfered at $Rx_1$ (and $pN_B$ $b$-symbols at $Rx_2$). At the end of a block, due to feedback from $Rx_1$, $Tx_1$ knows exactly which of its transmitted $a$-symbols caused interference at $Rx_{2}$ (since the same state variable $S_1[t]$ holds for both the receivers). For the next block, $Tx_1$ creates $N_B$ linear combinations of these $pN_B$ $a$-symbols (which caused interference at $Rx_2$ in the previous block) and sends these $N_B$ linear combinations as $c_2[t]$ (in the second subcarrier) over the next $N_B$ time slots. Due to bursty interference, w.h.p. only $pN_B$ of these linear combinations appear at $Rx_2$; but this is sufficient to decode $pN_B$ $a$-symbols constituting the linear combinations. Using these $a$-symbols $Rx_2$ can now recover all the interfered $b$-symbols in the previous block and hence achieve rate $1$ from the first subcarrier (same for $Rx_1$ due to symmetry). For the remaining levels in the second subcarrier, the following is done: lowest levels are not used ($c_3[t] = d_3[t]=0$), and the transmitters send fresh symbols in the highest level which appear interference free at the receivers (as the lowest levels are not used). This leads to an additional rate $1$ from the second subcarrier. Adding rates from the two subcarriers, we achieve symmetric rate $2$. This is in fact the symmetric capacity; an easy consequence of the outer bounds developed in this paper.

\par The above example demonstrates a \textit{helping} mechanism; the second subcarrier \textit{helped} the first subcarrier in recovering interfered symbols in a pipelined fashion. In this paper, we generalize this idea for an arbitrary collection of subcarriers with the following constraint: interference states across subcarriers are drawn from an arbitrary joint distribution (instantiated i.i.d. over time) and the marginal probability of interference is same for each subcarrier. The main idea behind the generalization is to use specific \textit{levels} in very strongly interfered subcarriers to recover interfered signals for strongly and weakly interfered subcarriers in a pipelined fashion as shown in the toy example. Another aspect captured by the toy example is the importance of burstiness; subcarriers in the above example are separable (due to our results and \cite{ihsiang_bursty_feedback}) when interference is always present. Hence, the proposed helping mechanism owes its relevance to bursty interference.
Our main contributions are as follows:
\begin{itemize}
\item In the linear deterministic setup, we have a complete capacity region characterization. In the setup with Gaussian noise, we have a tight generalized degrees of freedom (GDoF) characterization and provide outer bounds on the capacity region.
\item The inner and outer bounds are non-trivial extensions of single carrier results \cite{ihsiang_bursty_feedback}. We identify regimes where treating subcarriers separately is optimal. For the remaining regimes, we employ coding across subcarriers (helping mechanism) to achieve tight results. The outer bounds involve a subset entropy inequality by Madiman and Tetali \cite{madiman_tetali}.
\end{itemize}
The remainder of this paper is organized as follows. Section~\ref{sec:model} deals with the notation and setup. Section~\ref{sec:main_results} summarizes the main results of this paper. This is followed by Section~\ref{sec:outer_bounds} on outer bounds and Section~\ref{sec:inner_bounds} on inner bounds. We conclude the paper with Section~\ref{sec:gdof} on the GDoF characterization.
\section{Notation and Setup} \label{sec:model}
We consider a system with two base stations (transmitters) $Tx_1$ and $Tx_2$, and two users (receivers) $Rx_1$ and $Rx_2$. For $i\in \{1,2\}$, $Tx_i$ has message $W^{(i)}$ for $Rx_i$. There are $M$ parallel channels from $Tx_i$ to $Rx_i$ (subcarriers indexed by $j\in\{1,2,\ldots M \}$). 
In this paper, we consider two setups for the subcarrier channel: the first one is based on the linear deterministic model \cite{ADT_det_channel} (LD setup), and the second one is based on the Gaussian interference channel (GN setup). The subcarrier channel model for both the setups, followed by the statistics of bursty interference and rate requirements are described below.

\begin{figure}[!ht]
\begin{center}
\includegraphics[scale=0.51]{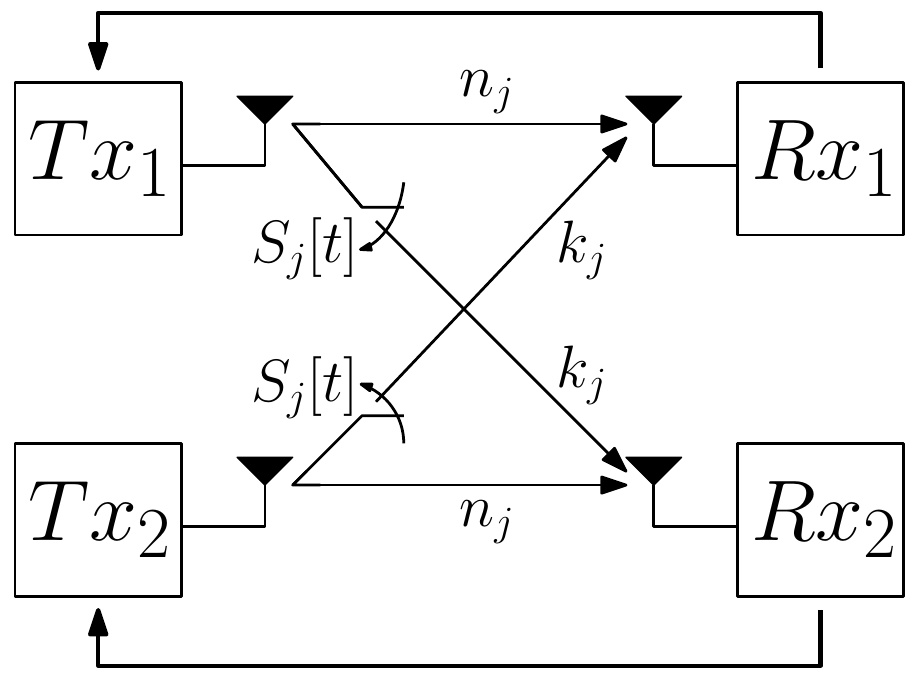}
\caption{Bursty interference channel (with feedback) for subcarrier $j$ in LD setup: $n_j$ and $k_j$ represent direct and interfering link strengths. Presence of interference at time index $t$ is determined by Bernoulli random variable $S_j[t]$.}
\label{fig:system_model}
\end{center}
\end{figure}

\paragraph*{Subcarrier channel model} In the LD setup, each subcarrier is modeled by a $2$-user (symmetric) LDIC \cite{ADT_det_channel} with a bursty interfering link (explained below) and feedback from respective receivers. At discrete time index $t \in \{1,2,\ldots N\}$, the transmitted signal in subcarrier $j$ of $Tx_i$ is $\mathbf{x}^{(i)}_j[t] \in \mathbb{F}^{q_j}$ where $\mathbb{F}$ is a finite field. The received signal in subcarrier $j$ at $Rx_i$ is given by:
\begin{align}
\mathbf{y}^{(i)}_j[t] = \mathbf{G}_j^{q_j-n_j}\mathbf{x}^{(i)}_j[t] + S_j[t] \mathbf{G}_j^{q_j-k_j}\mathbf{x}^{(i')}_j[t]  
\end{align}
where $\mathbf{G}_j$ is a $q_j \times q_j$ shift matrix in the terminology of deterministic channel models \cite{ADT_det_channel}, $S_j[t]$ is a Bernoulli random variable (details in interference statistics below) determining the presence of interference in subcarrier $j$ at time index $t$, $\mathbf{x}^{(i')}_j[t]$ denotes the transmitted signal on subcarrier $j$ of user $i' \neq i$, and parameters $n_j$ and $k_j$ represent the direct and interfering link strengths \cite{ADT_det_channel} in subcarrier $j$. Figure~\ref{fig:system_model} shows the channel model for subcarrier $j$. Without loss of generality, we assume $q_j=\max(n_j,k_j)$ and let $\alpha_j = \frac{k_j}{n_j}$ denote the normalized strength of the interfering signal in subcarrier $j$. For every time instant, it is convenient to consider a subcarrier as indexed levels of bit pipes \cite{ADT_det_channel}; each bit pipe carries a symbol from $\mathbb{F}$.
\par In the GN setup, at discrete time index $t \in \{1,2,\ldots N\}$, the transmitted signal in subcarrier $j$ of $Tx_i$ is $x^{(i)}_j[t] \in \mathbb{C}$, such that $\frac{1}{N} \sum_{t=1}^N |x^{(i)}_j[t]|^2 \leq 1$. The received signal in subcarrier $j$ at $Rx_i$ is given by:
\begin{align}
y^{(i)}_j[t] = g_{D,j} x^{(i)}_j[t] + S_j[t]  g_{I,j} x^{(i')}_j[t] + z^{(i)}_j[t]  
\end{align}
where $g_{D,j}, \; g_{I,j} \in \mathbb{C}$ denote the direct and interfering channel gains, and $z^{(i)}_j[t] \sim \mathcal{CN}(0,1)$ is Gaussian noise. As in the LD setup, $S_j[t]$ is the interference state. In both LD and GN setups, $Tx_i$ receives causal feedback from $Rx_i$ (feedback consists of the received signal and the interference state).
\paragraph*{Interference statistics} We consider the same interference statistics for both LD and GN setups. As described above, the presence of interference in subcarrier $j$ at time index $t$ is given by a Bernoulli random variable $S_j[t]$ (takes values in $\{0,1\}$). The $M$ Bernoulli random variables $\{S_1[t], S_2[t],\ldots S_M[t]\}$ have a joint probability distribution $\mathbb{P}(S_1[t]=s_1,S_2[t]=s_2, \ldots S_M[t] = s_M)=\mathbb{P}(S_1=s_1,S_2=s_2, \ldots S_M = s_M) $ instantiated i.i.d. over time. In this paper, we restrict the analysis to joint distributions with the same marginal probabilities for every $S_j[t]$ \emph{i.e.,} $\forall j,\; \mathbb{E}(S_j[t]) = p$. The transmitters are assumed to know the above statistics, but are limited to causal information on the interference realizations in the subcarriers (through feedback).
\paragraph*{Rate requirements} We consider the same rate requirements for both LD and GN setups. Base station $Tx_i$ intends to send message $W^{(i)}$ to $Rx_i$ over $N$ time slots (time index $t \in \{1,2,\ldots N\}$). Rate $R^{(i)}$ (corresponding to $W^{(i)}$) is considered achievable if the probability of decoding error is vanishingly small as $N\rightarrow \infty$. 
\section{Main Results} \label{sec:main_results}
\begin{figure}[!ht]
\begin{center}
\includegraphics[scale=1.34]{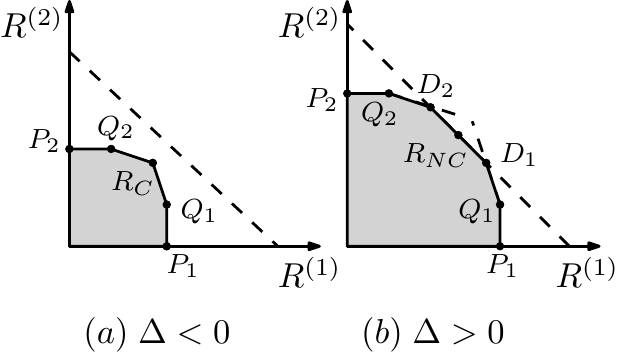}
\caption{Capacity region (LD setup) when $\Delta < 0$ and $\Delta > 0$. The dashed line representing inequality~(\ref{eq:OB_3}) is active only when $\Delta >0$. Symmetric capacity ($C_{sym}$) for $\Delta < 0$ and $\Delta \geq 0$ is given by $R_{C}= \frac{p}{1+p}\Delta   + \sum_{j=1}^{M}  n_j$ and $R_{NC} = \frac{p}{2}\Delta   + \sum_{j=1}^{M}  n_j $ respectively.
}
\label{fig:crossover}
\end{center}
\end{figure}

\begin{theorem}[LD setup capacity]
The capacity region for $(R^{(1)}, R^{(2)})$ in the LD setup is given by the following rate inequalities,
\begin{align}
R^{(i)}& \leq p\Delta +  \sum_{j=1}^M n_j(1+p) - (n_j-k_j)^{+}p \label{eq:OB_4} \\
R^{(i)} + p R^{(i')} & \leq  p \Delta  + \sum_{j=1}^{M}  n_j  (1+p)  \label{eq:causal_LD}\\
R^{(i)} + R^{(i')} &\leq  p\Delta   + 2\sum_{j=1}^{M}  n_j  \label{eq:OB_3} 
\end{align}
where $i,\;i' \in\{1,2\}$ and $i\neq i'$, and $\Delta =\displaystyle{\sum_{j=1}^{M}}  \max(n_j,k_j) + (n_j-k_j)^{+} - 2n_j = \sum_{j : \alpha_j > 2} (k_j  - 2 n_j ) -  \sum_{j : \alpha_j \leq 1 } k_j  - \sum_{j : 1 < \alpha_j \leq 2} (2n_j  - k_j ) $. 
\end{theorem}
As shown in Figure~\ref{fig:crossover}, the shape of the capacity region depends on the value of $\Delta$. An intuitive interpretation of $\Delta$ comes from our inner bounds; $\Delta > 0$ implies there are enough levels in subcarriers with $\alpha_j > 2$ (very strong interference) to recover the interfered signals for subcarriers with $\alpha_j \leq 1$ (weak interference) and $1 <\alpha_j \leq 2$ (strong interference). 
The details of the rate tuples $(R^{(1)},R^{(2)})$ marked in Figure~\ref{fig:crossover} are listed below:\\
$\bullet$ $P_1$ : $( p \Delta  + \sum_{j=1}^{M} n_j(1+p) - (n_j-k_j)^+ p , 0 )$ \\
$\bullet$ $Q_1 : ( p \Delta  + \sum_{j=1}^{M} n_j(1+p) - (n_j-k_j)^+p ,\sum_{j=1}^M(n_j-k_j)^+ ) $\\
$\bullet$ $D_1 : (p\Delta   + \sum_{j=1}^{M}  n_j , \sum_{j=1}^{M}  n_j )$\\
$\bullet$ $R_{C} \equiv (R_{C}, R_{C})  : (\frac{p}{1+p}\Delta   + \sum_{j=1}^{M}  n_j , \frac{p}{1+p}\Delta   + \sum_{j=1}^{M}  n_j )$\\
$\bullet$ $R_{NC} \equiv (R_{NC}, R_{NC}) : (\frac{p}{2}\Delta   + \sum_{j=1}^{M}  n_j , \frac{p}{2}\Delta   + \sum_{j=1}^{M}  n_j )$\\
$\bullet$ $D_2 : ( \sum_{j=1}^{M}  n_j , p \Delta   + \sum_{j=1}^{M}  n_j  )$\\
$\bullet$ $Q_2 : ( \sum_{j=1}^{M} (n_j-k_j)^+ , p \Delta  + \sum_{j=1}^{M} n_j(1+p) -(n_j-k_j)^+p) $\\
$\bullet$ $P_2$ : $(0, p \Delta  + \sum_{j=1}^{M} n_j(1+p) -(n_j-k_j)^+p )$. 
\begin{theorem}[GN setup outer bounds]
The following rate inequalities are outer bounds on achievable $(R^{(1)}, R^{(2)})$ in the GN setup.
\begin{align}
R^{(i)} &\leq \sum_{j=1}^M  (1-p) \log \left(  1  + |g_{D,j}|^2 \right) + p \log \left( 1+ |g_{D,j}|^2 +|g_{I,j}|^2   \right ) \label{eq:OB_R^i_GN} \\
R^{(i)} + p R^{(i')} & \leq p  \Delta_{G} + (1+p) \sum_{j=1}^{M} \log \left ( 1+|g_{D,j}|^2   \right )  \label{eq:causal_GN} \\
R^{(i)} + R^{(i')} & \leq   p \Delta_{G} + 2 \sum_{j=1}^{M} \log \left ( 1+|g_{D,j}|^2   \right ) \label{eq:OB_sum_GN}
\end{align}
where $i,\;i' \in\{1,2\}$ and $i\neq i'$, and $\Delta_G =  \sum_{j=1}^M  \log   \left ( 1+ \left ( |g_{D,j}| + |g_{I,j}|  \right )^2  \right)  +  \log \left (  1+ \frac{|g_{D,j}|^2}{1+ |g_{I,j}|^2} \right ) -2 \log \left ( 1+|g_{D,j}|^2   \right ) $.
\end{theorem}
\begin{theorem}[GN setup GDoF]
In the GN setup, assuming $g_{D,j} = \sqrt{SNR}$,  $g_{I,j} = \sqrt{INR_j}$ and $INR_j = SNR^{\beta_j}$ (rational $\beta_j$),
\begin{align}
GDoF\left( \beta_1,\ldots \beta_M \right) & = \limsup_{SNR \rightarrow \infty} \frac{ C_{sym} \left( SNR, \beta_1, \ldots \beta_M \right) }{M \log(SNR)} \nonumber \\
&= 1+ \min \left( \frac{\frac{p}{2} \Delta_{GDoF}}{M} ,  \frac{\frac{p}{1+p} \Delta_{GDoF}}{M} \right) \label{eq:GDoF}
\end{align}
where $C_{sym}$ denotes the symmetric capacity and $\Delta_{GDoF} = \sum_{j=1}^M ( \max \left( 1, \beta_j \right)  +  (1-\beta_j)^{+}  - 2)$. 
\end{theorem}
\begin{corollary}[separability] \label{cor:separate}
In the LD setup, for achieving symmetric capacity, treating subcarriers separately is optimal when all $\alpha_j \leq 2$ or all $\alpha_j \geq 2$ (and for the degenerate case of $p\in\{1,0\}$). For the remaining cases, coding across subcarriers achieves symmetric capacity. Similarly, in the GN setup, treating subcarriers separately is GDoF optimal when all $\beta_j \leq 2$ or all $\beta_j \geq 2$.
\end{corollary}

\section{Outer bounds: LD and GN setups} \label{sec:outer_bounds}
In this section, we focus on proofs of outer bounds in the LD and GN setups. We refer to outer bounds (\ref{eq:causal_LD}) and (\ref{eq:causal_GN}) as causal outer bounds as they account for the causal knowledge of subcarrier interference states at the transmitter\footnote{In Figure~\ref{fig:crossover}, for $\Delta <0$, the symmetric capacity $R_{C}$ stems from causal outer bound (\ref{eq:causal_LD}); hence the subscript ``C'' for causal.}. For proving these causal outer bounds, we use a subset entropy inequality by Madiman and Tetali which we describe in Section~\ref{sec:madiman_tetali}, prior to the proofs. Then we introduce some additional notation in Section~\ref{sec:add_notation} followed by outer bound proofs for the LD setup (Section~\ref{sec:LD_OB}) and GN setup (Section~\ref{sec:GN_OB}). 
\subsection{Madiman-Tetali subset inequality} \label{sec:madiman_tetali}
We now describe a subset entropy inequality by Madiman and Tetali \cite{madiman_tetali}. Consider a hypergraph $(U, \mathcal{E})$ where $U$ is a finite ground set and $\mathcal{E}$ is a collection of subsets of $U$. A function $\mathcal{G} : \mathcal{E} \rightarrow \mathbb{R}^+$ is called a fractional partition of $(U, \mathcal{E})$ if it satisfies the following condition $\forall j \in U$.
\begin{align}
\sum_{E \in \mathcal{E} : j \in E } \mathcal{G}(E) = 1 \label{eq:fractional_cover}
\end{align} 
With the above definition, the subset entropy inequality can now be stated as follows,
\begin{align}
\sum_{E \in \mathcal{E} } \mathcal{G}(E) H(X_E) \geq H(X_U) \label{eq:madiman_tetali}
\end{align}
where $\mathcal{G}$ is a fractional partition and the above inequality holds for any collection of jointly distributed random variables $X_U$. The differential entropy version of the above inequality has the same form \cite{madiman_tetali}. To use these inequalities in our setups, we first choose a suitable fractional partition as explained below. For $\mathbf{s}\in \{0,1\}^M$, let $\mathbf{S}[t]= \left (S_1[t],\; S_2[t],\ldots S_M[t] \right) = \mathbf{s}$ denote the collection of interference states of all the $M$ subcarriers at time index $t$. As specified in Section~\ref{sec:model}, the occurrence of $\mathbf{S}[t]=\mathbf{s}$ is governed by the joint probability distribution $\mathbb{P}(\mathbf{S}[t]=\mathbf{s})$. To define a fractional partition, we consider the ground set $U=\{1,2,\ldots M\}$ (\emph{i.e.,} the index set of subcarriers) and view $\mathbf{s} \in \{0,1\}^M$ as a collection of $M$ indicator functions for representing any subset of $U$. The power set of $U$ (excluding subsets $\mathbf{s}$ such that $\mathbb{P}(\mathbf{S}[t]=\mathbf{s})=0$) is chosen as set $\mathcal{E}$. Now, we define a fractional partition $\mathcal{G}: \mathcal{E} \rightarrow \mathbb{R}^+$ as follows.
\begin{align}
\mathcal{G}(E) = \frac{ \mathbb{P}(\mathbf{S}[t]=\mathbf{s}_E) }{p} \label{eq:fractional_cover_setup}
\end{align}
where $E\in \mathcal{E}$ and $\mathbf{s}_E$ denotes the joint state where only the subcarriers whose index is in set $E$ face interference. The fractional partition condition holds as follows. 
\begin{align}
\sum_{E \in \mathcal{E} : j \in E } \frac{ \mathbb{P}(\mathbf{S}[t]=\mathbf{s}_E) }{p} = \frac{\mathbb{E}(S_j[t])}{p}=1
\end{align}
In Section~\ref{sec:LD_causal}, we demonstrate the application of inequality (\ref{eq:madiman_tetali}), in conjunction with the fractional partition defined in (\ref{eq:fractional_cover_setup}), for proving outer bound (\ref{eq:causal_LD}). Similarly, in Section~\ref{sec:causal_GN}, for proving outer bound (\ref{eq:causal_GN}) we use the differential entropy version \cite{madiman_tetali} of inequality (\ref{eq:madiman_tetali}) with the same fractional partition.
\subsection{Additional notation} \label{sec:add_notation}
For notational convenience, we use indicator functions $\mathbb{I}_{j \not \in \mathbf{s}}$ and $\mathbb{I}_{j \in \mathbf{s}}$ to denote the absence and presence of interference in subcarrier $j$ when the joint state realization across $M$ subcarriers is $\mathbf{S}[t]=\mathbf{s} \in \{0,1\}^M$.
Also, in the proofs we use $\displaystyle{\sum_{\mathbf{s}}}$ to denote $\displaystyle{\sum_{\mathbf{s} \in \{0,1\}^M}}$.
For convenience, we have listed all the additional notation used for outer bound proofs in LD and GN setups (some notation is common to both setups).
\par Notation used in LD setup proofs:
\begin{itemize}
\item $\mathbf{S}_{1:t} = (\mathbf{S}[1],\;\mathbf{S}[2],\;\ldots \mathbf{S}[t] )$.

\item $\mathbf{Y}^{(i)}[t] =(\mathbf{y}^{(i)}_1[t],\;\mathbf{y}^{(i)}_2[t],\ldots \mathbf{y}^{(i)}_M[t])$ \emph{i.e.,} received signal (across $M$ subcarriers) for $Rx_i$ at time index $t$.

\item $\mathbf{Y}_{\mathbf{s}}^{(i)}[t]$: received signal (across $M$ subcarriers) for $Rx_i$ at time $t$ when $\mathbf{S}[t]=\mathbf{s}$. The difference between $\mathbf{Y}^{(i)}[t]$ and $\mathbf{Y}_{\mathbf{s}}^{(i)}[t]$ is that the state at time $t$ is assumed to be $\mathbf{s}$ in the latter.

\item $\mathbf{Y}^{(i)}_{1:t} = (\mathbf{Y}^{(i)}[1],\;\mathbf{Y}^{(i)}[2],\;\ldots\mathbf{Y}^{(i)}[t])$.

\item $\mathbf{V}_{\mathbf{s}}^{(i)}[t]$: interfering signals (across $M$ subcarriers) for $Rx_i$ when $\mathbf{S}[t]=\mathbf{s}$.

\item  $ \mathbf{V}^{(i)}_{1:t} =( \mathbf{V}_{\mathbf{S}[1]}^{(i)}[1],\; \mathbf{V}_{\mathbf{S}[2]}^{(i)}[2], \ldots \mathbf{V}_{\mathbf{S}[t]}^{(i)}[t])$.

\item $\tilde{\mathbf{V}}^{(i)}[t]$: interfering signals (across $M$ subcarriers) at $Rx_i$ when all its subcarriers face interference at time index $t$. This is equivalent to $\mathbf{V}_{\mathbf{s}}^{(i)}[t]$ with $\mathbf{s}=\{1,1,\ldots1\}$.

\item $\hat{\mathbf{X}}^{(i)}[t]$: received signal (across $M$ subcarriers) at $Rx_i$ when all its subcarriers are interference free at time index $t$.
This is equivalent to $\mathbf{Y}_{\mathbf{s}}^{(i)}[t]$ with $\mathbf{s}=\{0,0,\ldots0\}$.

\end{itemize}
\par Notation used in GN setup proofs:
\begin{itemize}
\item $\mathbf{S}_{1:t} = (\mathbf{S}[1],\;\mathbf{S}[2],\;\ldots \mathbf{S}[t] )$.

\item $\mathbf{Y}^{(i)}[t] =(y^{(i)}_1[t],\; y^{(i)}_2[t],\ldots y^{(i)}_M[t])$ \emph{i.e.,} received signal (across $M$ subcarriers) for $Rx_i$ at time index $t$.

\item $\mathbf{Y}_{\mathbf{s}}^{(i)}[t]$: received signal (across $M$ subcarriers) for $Rx_i$ at time $t$ when $\mathbf{S}[t]=\mathbf{s}$. The difference between $\mathbf{Y}^{(i)}[t]$ and $\mathbf{Y}_{\mathbf{s}}^{(i)}[t]$ is that the state at time $t$ is assumed to be $\mathbf{s}$ in the latter.

\item $\mathbf{Y}^{(i)}_{1:t} = (\mathbf{Y}^{(i)}[1],\;\mathbf{Y}^{(i)}[2],\;\ldots\mathbf{Y}^{(i)}[t])$.

\item $\mathbf{Z}^{(i)}[t]=(z^{(i)}_1[t],\; z^{(i)}_2[t],\ldots z^{(i)}_M[t])$ \emph{i.e.,} receiver noise (across $M$ subcarriers) for $Rx_i$ at time index $t$.

\item $\mathbf{Z}_{\mathbf{s}}^{(i)}[t]$: receiver noise in interfered subcarriers for $Rx_i$ at time index $t$ when $\mathbf{S}[t]=\mathbf{s}$.

\item $\mathbf{Z}^{(i)}_{\mathbf{s}^c}[t]$: receiver noise in interference free subcarriers for $Rx_i$ at time index $t$ when $\mathbf{S}[t]=\mathbf{s}$.

\item  $\mathbf{V}_{\mathbf{s}}^{(i)}[t] \oplus \mathbf{Z}^{(i)}[t]=( S_1[t]  g_{I,1} x^{(i')}_1[t] + z^{(i)}_1[t], \;S_2[t]  g_{I,2} x^{(i')}_2[t] + z^{(i)}_2[t], \ldots  S_M[t]  g_{I,M} x^{(i')}_M[t] + z^{(i)}_M[t]   )$ \emph{i.e.,} interfering signal (if present) plus noise, across $M$ subcarriers, for $Rx_i$ at time index $t$ when $\mathbf{S}[t]=\mathbf{s}$.

\item $\mathbf{V}_{\mathbf{s}}^{(i)}[t] \oplus \mathbf{Z}_{\mathbf{s}}^{(i)}[t]$: interfering signal plus noise in interfered subcarriers for $Rx_i$ at time index $t$ when $\mathbf{S}[t]=\mathbf{s}$. Note that this does not include the noise terms for subcarriers which do not face interference at time $t$ (unlike $\mathbf{V}_{\mathbf{s}}^{(i)}[t] \oplus \mathbf{Z}^{(i)}[t]$).

\item $\mathbf{V}^{(i)}_{1:t}\oplus \mathbf{Z}^{(i)}_{1:t} = (\mathbf{V}_{\mathbf{S}[1]}^{(i)}[1] \oplus \mathbf{Z}^{(i)}[1], \; \mathbf{V}_{\mathbf{S}[2]}^{(i)}[2] \oplus \mathbf{Z}^{(i)}[2] ,\ldots \mathbf{V}_{\mathbf{S}[t]}^{(i)}[t] \oplus \mathbf{Z}^{(i)}[t])$.

\item $\tilde{\mathbf{V}}^{(i)}[t] \oplus  \mathbf{Z}^{(i)}[t]$: interfering signal plus noise (across $M$ subcarriers) at $Rx_i$ when all its subcarriers face interference at time index $t$. This is equivalent to $\mathbf{V}_{\mathbf{s}}^{(i)}[t] \oplus \mathbf{Z}^{(i)}[t]$ with $\mathbf{s}=\{1,1,\ldots1\}$.

\item $\hat{\mathbf{X}}^{(i)}[t]\oplus \mathbf{Z}^{(i)}[t]$: received signal (across $M$ subcarriers) at $Rx_i$ when all its subcarriers are interference free at time index $t$. This is equivalent to $\mathbf{Y}_{\mathbf{s}}^{(i)}[t]$ with $\mathbf{s}=\{0,0,\ldots0\}$.

\item $\hat{\mathbf{X}}^{(i)}_{1:t} \oplus \mathbf{Z}^{(i)}_{1:t} = (\hat{\mathbf{X}}^{(i)}[1]\oplus \mathbf{Z}^{(i)}[1],\;\hat{\mathbf{X}}^{(i)}[2]\oplus \mathbf{Z}^{(i)}[2],\ldots \hat{\mathbf{X}}^{(i)}[t]\oplus \mathbf{Z}^{(i)}[t])$.
\end{itemize}

\subsection{Outer bounds: LD setup} \label{sec:LD_OB}
\subsubsection{Proof of outer bound (\ref{eq:OB_4})} See Appendix~\ref{sec:LD_OB_R1}.
\subsubsection{Proof of outer bound (\ref{eq:causal_LD})} \label{sec:LD_causal}
Using Fano's inequality for $Rx_1$, for any $\epsilon > 0$, there exists a large enough $N$ such that;
\begin{align}
&N R^{(1)} - N \epsilon \nonumber\\
& \leq I(W^{(1)} ; \mathbf{Y}^{(1)}_{1:N}, \mathbf{S}_{1:N}) \nonumber \\
&= \sum_{t=1}^{N}  I(W^{(1)};  \mathbf{Y}^{(1)}[t] | \mathbf{Y}^{(1)}_{1:t-1} ,\mathbf{S}_{1:t-1}, \mathbf{S}[t] ) \nonumber\\
&= \sum_{t=1}^{N} \sum_{\mathbf{s} } \mathbb{P}(\mathbf{S}[t]=\mathbf{s}) H( \mathbf{Y}_{\mathbf{s}}^{(1)}[t] | \mathbf{Y}^{(1)}_{1:t-1}, \mathbf{S}_{1:t-1} ,\mathbf{S}[t]=\mathbf{s} ) \nonumber \\ &\quad  - \sum_{t=1}^{N} \sum_{\mathbf{s}} \mathbb{P}(\mathbf{S}[t]=\mathbf{s}) H( \mathbf{Y}_{\mathbf{s}}^{(1)}[t] | W^{(1)}, \mathbf{Y}^{(1)}_{1:t-1}, \mathbf{S}_{1:t-1} , \mathbf{S}[t]= \mathbf{s} ) \nonumber\\
&\leq \sum_{t=1}^{N}  \sum_{\mathbf{s} }  \mathbb{P}(\mathbf{S}[t]=\mathbf{s}) \sum_{j=1}^{M} n_j \mathbb{I}_{j \not \in \mathbf{s}} + \max(n_j,k_j) \mathbb{I}_{j \in \mathbf{s}}  \nonumber \\ &\quad - \sum_{t=1}^{N}  \sum_{\mathbf{s}}  \mathbb{P}(\mathbf{S}[t]=\mathbf{s})  H( \mathbf{V}_{\mathbf{s}}^{(1)}[t] | W^{(1)}, \mathbf{Y}^{(1)}_{1:t-1} , \mathbf{S}_{1:t-1} ,\mathbf{S}[t]=\mathbf{s}) \nonumber\\
&= \sum_{t=1}^{N}  \sum_{j=1}^{M}  \sum_{ \mathbf{s}}  n_j \mathbb{P}(\mathbf{S}[t]=\mathbf{s})  \mathbb{I}_{j \not \in \mathbf{s}} + \max(n_j,k_j)  \mathbb{P}(\mathbf{S}[t]=\mathbf{s}) \mathbb{I}_{j \in \mathbf{s}}  \nonumber\\ & \quad - \sum_{t=1}^{N}  \sum_{\mathbf{s}}  \mathbb{P}(\mathbf{S}[t]=\mathbf{s})  H( \mathbf{V}_{\mathbf{s}}^{(1)}[t] | W^{(1)}, \mathbf{Y}^{(1)}_{1:t-1}, \mathbf{S}_{1:t-1}, \mathbf{S}[t]=\mathbf{s}) \nonumber\\
&= N \sum_{j=1}^{M}  (1-p) n_j + p\max(n_j,k_j)   \nonumber \\ &\quad- \sum_{t=1}^{N}  \sum_{\mathbf{s}}  \mathbb{P}(\mathbf{S}[t]=\mathbf{s})  H( \mathbf{V}_{\mathbf{s}}^{(1)}[t] | W^{(1)}, \mathbf{Y}^{(1)}_{1:t-1} ,\mathbf{S}_{1:t-1} ,\mathbf{S}[t]=\mathbf{s}) \nonumber
\end{align}
\begin{align}
&=N  \sum_{j=1}^{M} (1-p) n_j + p\max(n_j,k_j)    -  \sum_{t=1}^{N}  \sum_{\mathbf{s}}  \mathbb{P}(\mathbf{S}[t]=\mathbf{s})  H( \mathbf{V}_{\mathbf{s}}^{(1)}[t] | W^{(1)},  \mathbf{V}^{(1)}_{1:t-1}, \mathbf{S}_{1:t-1}) \nonumber\\
&\stackrel{(a)}\leq N  \sum_{j=1}^{M} (1-p) n_j + p\max(n_j,k_j)    - p \sum_{t=1}^{N}  H( \tilde{\mathbf{V}}^{(1)}[t] | W^{(1)},  \mathbf{V}^{(1)}_{1:t-1}, \mathbf{S}_{1:t-1}) \nonumber\\
&= N  \sum_{j=1}^{M} n_j  +  (\max(n_j,k_j)-n_j) p   -  p \sum_{t=1}^{N}  H( \tilde{\mathbf{V}}^{(1)}[t] | W^{(1)},  \mathbf{V}^{(1)}_{1:t-1}, \mathbf{S}_{1:t-1}) \label{eq:causal_1}
\end{align}
where (a) follows by using (\ref{eq:madiman_tetali}) for the fractional partition defined in (\ref{eq:fractional_cover_setup}).
\par Using Fano's inequality for $Rx_2$, for any $\epsilon > 0$, there exists a large enough $N$ such that;
\begin{align}
& N R^{(2)} - N \epsilon \nonumber\\
& \leq I(W^{(2)} ; \mathbf{Y}^{(1)}_{1:N}  , \mathbf{Y}^{(2)}_{1:N}, \mathbf{S}_{1:N} ,W^{(1)} ) \nonumber \\
& = I(W^{(2)} ; \mathbf{Y}^{(1)}_{1:N} ,  \mathbf{Y}^{(2)}_{1:N} ,\mathbf{S}_{1:N} | W^{(1)} ) \nonumber \\
& = \sum_{t=1}^{N}  I(W^{(2)}; \mathbf{Y}^{(2)}[t] , \mathbf{Y}^{(1)}[t] |  \mathbf{Y}^{(2)}_{1:t-1} ,  \mathbf{Y}^{(1)}_{1:t-1}  ,\mathbf{S}_{1:t-1} , W^{(1)} , \mathbf{S}[t]  ) \nonumber\\
& = \sum_{t=1}^{N} \sum_{ \mathbf{s}  }   \mathbb{P}(\mathbf{S}[t]=\mathbf{s})( H ( \mathbf{Y}_{\mathbf{s}}^{(2)}[t],\mathbf{Y}_{\mathbf{s}}^{(1)}[t] | \mathbf{Y}^{(2)}_{1:t-1}, \mathbf{ Y}^{(1)}_{1:t-1} ,\mathbf{S}_{1:t-1} ,W^{(1)}, \mathbf{S}[t]=\mathbf{s})
\nonumber\\ &\quad 
-    H ( \mathbf{Y}_{\mathbf{s}}^{(2)}[t], \mathbf{Y}_{\mathbf{s}}^{(1)}[t] | \mathbf{Y}^{(2)}_{1:t-1},  \mathbf{Y}^{(1)}_{1:t-1} ,\mathbf{S}_{1:t-1}, W^{(1)} , W^{(2)} ,\mathbf{S}[t]= \mathbf{s} ) )\nonumber\\
& = \sum_{t=1}^{N} \sum_{\mathbf{s} }   \mathbb{P}(\mathbf{S}[t]=\mathbf{s})     H ( \mathbf{Y}_{\mathbf{s}}^{(2)}[t], \mathbf{Y}_{\mathbf{s}}^{(1)}[t] | \mathbf{Y}^{(2)}_{1:t-1} , \mathbf{Y}^{(1)}_{1:t-1} ,\mathbf{S}_{1:t-1}, W^{(1)}, \mathbf{S}[t]=\mathbf{s}   )\nonumber\\
& = \sum_{t=1}^{N} \sum_{\mathbf{s}}   \mathbb{P}(\mathbf{S}[t]=\mathbf{s})   H( \hat{\mathbf{X}}^{(2)}[t], \mathbf{V}_{\mathbf{s}}^{(1)}[t] | \mathbf{Y}^{(2)}_{1:t-1},  \mathbf{Y}^{(1)}_{1:t-1} ,\mathbf{S}_{1:t-1}, W^{(1)} ,\mathbf{S}[t]=\mathbf{s}  )\nonumber\\
& \leq  \sum_{t=1}^{N}  \sum_{\mathbf{s}}   \mathbb{P}(\mathbf{S}[t]=\mathbf{s})    H ( \hat{\mathbf{X}}^{(2)}[t]  , \tilde{\mathbf{V}}^{(1)}[t]  | \mathbf{Y}^{(2)}_{1:t-1},  \mathbf{Y}^{(1)}_{1:t-1}, \mathbf{S}_{1:t-1}, W^{(1)}, \mathbf{S}[t]=\mathbf{s})\nonumber\\
& = \sum_{t=1}^{N}   H (  \hat{\mathbf{X}}^{(2)}[t]  ,\tilde{\mathbf{V}}^{(1)}[t]  | \mathbf{Y}^{(2)}_{1:t-1},  \mathbf{Y}^{(1)}_{1:t-1}, \mathbf{S}_{1:t-1}, W^{(1)}  ) \sum_{\mathbf{s} }   \mathbb{P}(\mathbf{S}[t]=\mathbf{s}) \nonumber\\
& = \sum_{t=1}^{N}   H (  \hat{\mathbf{X}}^{(2)}[t] , \tilde{\mathbf{V}}^{(1)}[t]   | \mathbf{Y}^{(2)}_{1:t-1},   \mathbf{Y}^{(1)}_{1:t-1} , \mathbf{S}_{1:t-1} , W^{(1)}   )  \nonumber\\
& \leq \sum_{t=1}^{N}   H (  \hat{\mathbf{X}}^{(2)}[t] ,  \tilde{\mathbf{V}}^{(1)}[t]   |  \mathbf{Y}^{(1)}_{1:t-1} , \mathbf{S}_{1:t-1} , W^{(1)}   )  \nonumber\\
& =\sum_{t=1}^{N}   H ( \hat{\mathbf{X}}^{(2)}[t]  , \tilde{\mathbf{V}}^{(1)}[t]  | \mathbf{V}^{(1)}_{1:t-1}, \mathbf{S}_{1:t-1} , W^{(1)}     )  \label{eq:causal_2}
\end{align}
\par Using inequalities~(\ref{eq:causal_1}) and (\ref{eq:causal_2}),
\begin{align}
& N R^{(1)}-  N \epsilon + pN R^{(2)} - pN\epsilon \nonumber \\
& \leq N  \sum_{j=1}^{M} n_j  +  (\max(n_j,k_j)-n_j) p    + p  \sum_{t=1}^{N}   H (  \hat{\mathbf{X}}^{(2)}[t]  | \tilde{\mathbf{V}}^{(1)}[t],   W^{(1)},  \mathbf{V}^{(1)}_{1:t-1} ,\mathbf{S}_{1:t-1} ) \nonumber\\
& \leq N  \sum_{j=1}^{M} n_j  +  (\max(n_j,k_j)-n_j) p      +  p  \sum_{t=1}^{N}   H ( \hat{\mathbf{X}}^{(2)}[t]  | \tilde{\mathbf{V}}^{(1)}[t] ) \nonumber\\
& \leq   N  \sum_{j=1}^{M} n_j  +  (\max(n_j,k_j)-n_j) p  + p  \sum_{t=1}^{N}  \sum_{j=1}^{M} (n_j-k_j)^{+}\nonumber\\
&= N p \Delta +  N  \sum_{j=1}^{M} (1+p) n_j   
\end{align}
where $\Delta =  \sum_{j=1}^{M} \max(n_j,k_j)  + (n_j-k_j)^{+} - 2n_j $. The bound on $p R^{(1)} + R^{(2)}$ follows by symmetry, and this completes the proof of outer bound (\ref{eq:causal_LD}).

\par The above proof demonstrates a connection between subset entropy inequalities and bursty interference in multicarrier systems. In \cite{self_isit13_w}, we demonstrated a similar connection by using a sliding window subset entropy inequality \cite{Tie_liu} to show tight outer bounds for the case without feedback (in multicarrier systems with bursty interference).\\
\subsubsection{Proof of outer bound (\ref{eq:OB_3})} See Appendix~\ref{sec:LD_OB_R1_R2_sum}.

\subsection{Outer bounds: GN setup} \label{sec:GN_OB}
\subsubsection{Proof of outer bound (\ref{eq:OB_R^i_GN})} See Appendix \ref{sec:GN_R1_OB}.
\subsubsection{Proof of outer bound (\ref{eq:causal_GN})} \label{sec:causal_GN}
Using Fano's inequality for $Rx_1$, for any $\epsilon > 0$, there exists a large enough $N$ such that;
\begin{align}
&N R^{(1)} - N \epsilon \nonumber\\
& \leq I(W^{(1)} ; \mathbf{Y}^{(1)}_{1:N} ,\mathbf{S}_{1:N}) \nonumber \\
&= \sum_{t=1}^{N}  I(W^{(1)};  \mathbf{Y}^{(1)}[t] | \mathbf{Y}^{(1)}_{1:t-1}, \mathbf{S}_{1:t-1}, \mathbf{S}[t] ) \nonumber\\
&= \sum_{t=1}^{N} \sum_{\mathbf{s} } \mathbb{P}(\mathbf{S}[t]=\mathbf{s}) h( \mathbf{Y}^{(1)}[t] | \mathbf{Y}^{(1)}_{1:t-1} ,\mathbf{S}_{1:t-1} ,\mathbf{S}[t]=\mathbf{s} ) - \sum_{t=1}^{N} \sum_{\mathbf{s}} \mathbb{P}(\mathbf{S}[t]=\mathbf{s}) h( \mathbf{Y}^{(1)}[t] | W^{(1)} ,\mathbf{Y}^{(1)}_{1:t-1} ,\mathbf{S}_{1:t-1} , \mathbf{S}[t]= \mathbf{s} ) \nonumber\\
&\leq \sum_{t=1}^{N} \sum_{\mathbf{s} } \mathbb{P}(\mathbf{S}[t]=\mathbf{s}) h( \mathbf{Y}^{(1)}[t] |\mathbf{S}[t]=\mathbf{s} )  - \sum_{t=1}^{N} \sum_{\mathbf{s}} \mathbb{P}(\mathbf{S}[t]=\mathbf{s}) h( \mathbf{Y}^{(1)}[t] | W^{(1)}, \mathbf{Y}^{(1)}_{1:t-1} ,\mathbf{S}_{1:t-1} , \mathbf{S}[t]= \mathbf{s} ) \nonumber\\
&\stackrel{(a)}\leq   N  \sum_{j=1}^{M} (1-p) \log \left ( 1+|g_{D,j}|^2   \right )  + p \log   \left ( 1+ \left ( |g_{D,j}| + |g_{I,j}|  \right )^2  \right) + NM \log (\pi e) \nonumber\\
& \quad - \sum_{t=1}^{N} \sum_{\mathbf{s}} \mathbb{P}(\mathbf{S}[t]=\mathbf{s}) h( \mathbf{Y}^{(1)}[t] | W^{(1)} ,\mathbf{Y}^{(1)}_{1:t-1} ,\mathbf{S}_{1:t-1}  ,\mathbf{S}[t]= \mathbf{s} ) \nonumber\\
&= N  \sum_{j=1}^{M} (1-p) \log \left ( 1+|g_{D,j}|^2   \right )  + p \log   \left ( 1+ \left ( |g_{D,j}| + |g_{I,j}|  \right )^2  \right) + NM \log (\pi e) \nonumber\\ &\quad  - \sum_{t=1}^{N}  \sum_{\mathbf{s}}  \mathbb{P}(\mathbf{S}[t]=\mathbf{s})  h( \mathbf{V}_{\mathbf{s}}^{(1)}[t] \oplus \mathbf{Z}^{(1)}[t]  | W^{(1)}, \mathbf{Y}^{(1)}_{1:t-1}, \mathbf{S}_{1:t-1} ,\mathbf{S}[t]=\mathbf{s}) \nonumber\\
&= N  \sum_{j=1}^{M} (1-p) \log \left ( 1+|g_{D,j}|^2   \right )  + p \log   \left ( 1+ \left ( |g_{D,j}| + |g_{I,j}|  \right )^2  \right) + NM \log (\pi e) \nonumber\\  &\quad  - \sum_{t=1}^{N}  \sum_{\mathbf{s}}  \mathbb{P}(\mathbf{S}[t]=\mathbf{s})  h( \mathbf{V}_{\mathbf{s}}^{(1)}[t] \oplus \mathbf{Z}_{\mathbf{s}}^{(1)}[t]  | W^{(1)} ,\mathbf{Y}^{(1)}_{1:t-1} ,\mathbf{S}_{1:t-1}, \mathbf{S}[t]=\mathbf{s}) \nonumber\\ & \quad - \sum_{t=1}^{N}  \sum_{\mathbf{s}}  \mathbb{P}(\mathbf{S}[t]=\mathbf{s})  h( \mathbf{Z}_{\mathbf{s}^c}^{(1)}[t]  | \mathbf{V}_{\mathbf{s}}^{(1)}[t] \oplus \mathbf{Z}_{\mathbf{s}}^{(1)}[t] , W^{(1)} , \mathbf{Y}^{(1)}_{1:t-1} , \mathbf{S}_{1:t-1} ,\mathbf{S}[t]=\mathbf{s}) \nonumber\\
&= N  \sum_{j=1}^{M} (1-p) \log \left ( 1+|g_{D,j}|^2   \right )  + p \log   \left ( 1+ \left ( |g_{D,j}| + |g_{I,j}|  \right )^2  \right) + NM \log (\pi e) \nonumber\\  &\quad  - \sum_{t=1}^{N}  \sum_{\mathbf{s}}  \mathbb{P}(\mathbf{S}[t]=\mathbf{s})  h( \mathbf{V}_{\mathbf{s}}^{(1)}[t] \oplus \mathbf{Z}_{\mathbf{s}}^{(1)}[t]  | W^{(1)}, \mathbf{Y}^{(1)}_{1:t-1}, \mathbf{S}_{1:t-1} ,\mathbf{S}[t]=\mathbf{s})  - \sum_{t=1}^{N} \sum_{\mathbf{s}}  \mathbb{P}(\mathbf{S}[t]=\mathbf{s}) \sum_{j=1}^M  \mathbb{I}_{j \not \in \mathbf{s}}  \log(\pi e)  \nonumber \\
&=N  \sum_{j=1}^{M} (1-p) \log \left ( 1+|g_{D,j}|^2   \right )  + p \log   \left ( 1+ \left ( |g_{D,j}| + |g_{I,j}|  \right )^2  \right) + NM \log (\pi e) \nonumber\\ &\quad  - \sum_{t=1}^{N}  \sum_{\mathbf{s}}  \mathbb{P}(\mathbf{S}[t]=\mathbf{s})  h( \mathbf{V}_{\mathbf{s}}^{(1)}[t] \oplus \mathbf{Z}_{\mathbf{s}}^{(1)}[t]  | W^{(1)}, \mathbf{Y}^{(1)}_{1:t-1}, \mathbf{S}_{1:t-1} )  - N M (1-p)\log(\pi e)  \nonumber \\
&\stackrel{(b)}\leq N  \sum_{j=1}^{M} (1-p) \log \left ( 1+|g_{D,j}|^2   \right )  + p \log   \left ( 1+ \left ( |g_{D,j}| + |g_{I,j}|  \right )^2  \right) + NM \log (\pi e) \nonumber\\ &\quad  - p \sum_{t=1}^{N}  h( \tilde{\mathbf{V}}^{(1)}[t] \oplus  \mathbf{Z}^{(1)}[t] | W^{(1)} ,\mathbf{Y}^{(1)}_{1:t-1}, \mathbf{S}_{1:t-1}) - N M (1-p)\log(\pi e) \nonumber
\end{align}
\begin{align}
&=N  \sum_{j=1}^{M} (1-p) \log \left ( 1+|g_{D,j}|^2   \right )  + p \log   \left ( 1+ \left ( |g_{D,j}| + |g_{I,j}|  \right )^2  \right) + NM \log (\pi e) \nonumber\\ &\quad  - p \sum_{t=1}^{N}  h( \tilde{\mathbf{V}}^{(1)}[t] \oplus  \mathbf{Z}^{(1)}[t] | W^{(1)}, \mathbf{V}^{(1)}_{1:t-1}\oplus \mathbf{Z}^{(1)}_{1:t-1} , \mathbf{S}_{1:t-1}) - N M (1-p)\log(\pi e) \label{eq:GN_causal_a}
\end{align}
where (a) follows from the proof of (\ref{eq:GN_looser_R1}) (see Appendix~\ref{sec:GN_R1_OB}) and (b) follows by using the differential entropy version of inequality (\ref{eq:madiman_tetali}) for the fractional partition defined in (\ref{eq:fractional_cover_setup}).
\par Using Fano's inequality for $Rx_2$, for any $\epsilon > 0$, there exists a large enough $N$ such that;
\begin{align}
& N R^{(2)} - N \epsilon \nonumber\\
& \leq I(W^{(2)} ; \mathbf{Y}^{(1)}_{1:N}  , \mathbf{Y}^{(2)}_{1:N}, \mathbf{S}_{1:N} ,W^{(1)} ) \nonumber \\
& = I(W^{(2)} ; \mathbf{Y}^{(1)}_{1:N} ,  \mathbf{Y}^{(2)}_{1:N} ,\mathbf{S}_{1:N} | W^{(1)} ) \nonumber \\
& = \sum_{t=1}^{N}  I(W^{(2)}; \mathbf{Y}^{(2)}[t],  \mathbf{Y}^{(1)}[t] |  \mathbf{Y}^{(2)}_{1:t-1}  , \mathbf{Y}^{(1)}_{1:t-1}  ,\mathbf{S}_{1:t-1}, W^{(1)} , \mathbf{S}[t]  ) \nonumber\\
& = \sum_{t=1}^{N} \sum_{ \mathbf{s}  }   \mathbb{P}(\mathbf{S}[t]=\mathbf{s})  I(W^{(2)}; \mathbf{Y}^{(2)}[t]  ,\mathbf{Y}^{(1)}[t] |  \mathbf{Y}^{(2)}_{1:t-1},  \mathbf{Y}^{(1)}_{1:t-1} , \mathbf{S}_{1:t-1}, W^{(1)} , \mathbf{S}[t]=\mathbf{s}  ) \nonumber\\
& = \sum_{t=1}^{N} \sum_{ \mathbf{s}  }   \mathbb{P}(\mathbf{S}[t]=\mathbf{s})  I(W^{(2)}; \hat{\mathbf{X}}^{(2)}[t]\oplus \mathbf{Z}^{(2)}[t], \mathbf{V}^{(1)}_{\mathbf{s}}[t]  \oplus \mathbf{Z}^{(1)}[t] |  \mathbf{Y}^{(2)}_{1:t-1}   ,\mathbf{Y}^{(1)}_{1:t-1}  ,\mathbf{S}_{1:t-1} ,W^{(1)} , \mathbf{S}[t]=\mathbf{s}  ) \nonumber\\
& = \sum_{t=1}^{N} \sum_{ \mathbf{s}  }   \mathbb{P}(\mathbf{S}[t]=\mathbf{s})  I(W^{(2)}; \hat{\mathbf{X}}^{(2)}[t]\oplus \mathbf{Z}^{(2)}[t], \mathbf{V}_{\mathbf{s}}^{(1)}[t]  \oplus \mathbf{Z}^{(1)}_{\mathbf{s}}[t] |  \mathbf{Y}^{(2)}_{1:t-1} ,  \mathbf{Y}^{(1)}_{1:t-1}  ,\mathbf{S}_{1:t-1}, W^{(1)}  ,\mathbf{S}[t]=\mathbf{s}  ) \nonumber\\
&\quad + \sum_{t=1}^{N} \sum_{ \mathbf{s}  }   \mathbb{P}(\mathbf{S}[t]=\mathbf{s})  I(W^{(2)};  \mathbf{Z}^{(1)}_{\mathbf{s}^c}[t] | \hat{\mathbf{X}}^{(2)}[t]\oplus \mathbf{Z}^{(2)}[t], \mathbf{V}_{\mathbf{s}}^{(1)}[t]  \oplus \mathbf{Z}_{\mathbf{s}}^{(1)}[t] ,  \mathbf{Y}^{(2)}_{1:t-1}  , \mathbf{Y}^{(1)}_{1:t-1}  ,\mathbf{S}_{1:t-1}, W^{(1)}  ,\mathbf{S}[t]=\mathbf{s}  )  \nonumber\\
& = \sum_{t=1}^{N} \sum_{ \mathbf{s}  }   \mathbb{P}(\mathbf{S}[t]=\mathbf{s})  I(W^{(2)}; \hat{\mathbf{X}}^{(2)}[t]\oplus \mathbf{Z}^{(2)}[t], \mathbf{V}_{\mathbf{s}}^{(1)}[t]  \oplus \mathbf{Z}^{(1)}_{\mathbf{s}}[t] |  \mathbf{Y}^{(2)}_{1:t-1} ,  \mathbf{Y}^{(1)}_{1:t-1}  ,\mathbf{S}_{1:t-1}, W^{(1)}  ,\mathbf{S}[t]=\mathbf{s}  ) \nonumber\\
& \leq  \sum_{t=1}^{N} \sum_{ \mathbf{s}  }   \mathbb{P}(\mathbf{S}[t]=\mathbf{s})  I(W^{(2)}; \hat{\mathbf{X}}^{(2)}[t]\oplus \mathbf{Z}^{(2)}[t], \tilde{\mathbf{V}}^{(1)}[t] \oplus \mathbf{Z}^{(1)}[t] |  \mathbf{Y}^{(2)}_{1:t-1}  , \mathbf{Y}^{(1)}_{1:t-1} , \mathbf{S}_{1:t-1} ,W^{(1)} ,\mathbf{S}[t]=\mathbf{s}  ) \nonumber\\
& =  \sum_{t=1}^{N} \sum_{ \mathbf{s}  }   \mathbb{P}(\mathbf{S}[t]=\mathbf{s})  h(\hat{\mathbf{X}}^{(2)}[t]\oplus \mathbf{Z}^{(2)}[t], \tilde{\mathbf{V}}^{(1)}[t] \oplus \mathbf{Z}^{(1)}[t] |  \mathbf{Y}^{(2)}_{1:t-1}   ,\mathbf{Y}^{(1)}_{1:t-1} , \mathbf{S}_{1:t-1}, W^{(1)},  \mathbf{S}[t]=\mathbf{s}  ) \nonumber\\
& \quad -\sum_{t=1}^{N} \sum_{ \mathbf{s}  }   \mathbb{P}(\mathbf{S}[t]=\mathbf{s})  h(\hat{\mathbf{X}}^{(2)}[t]\oplus \mathbf{Z}^{(2)}[t], \tilde{\mathbf{V}}^{(1)}[t]  \oplus \mathbf{Z}^{(1)}[t] |  \mathbf{Y}^{(2)}_{1:t-1} ,  \mathbf{Y}^{(1)}_{1:t-1} , \mathbf{S}_{1:t-1}, W^{(1)}, W^{(2)} ,\mathbf{S}[t]=\mathbf{s}  ) \nonumber\\
& =  \sum_{t=1}^{N} \sum_{ \mathbf{s}  }   \mathbb{P}(\mathbf{S}[t]=\mathbf{s})  h(\hat{\mathbf{X}}^{(2)}[t]\oplus \mathbf{Z}^{(2)}[t], \tilde{\mathbf{V}}^{(1)}[t] \oplus \mathbf{Z}^{(1)}[t] |  \mathbf{Y}^{(2)}_{1:t-1} ,  \mathbf{Y}^{(1)}_{1:t-1} , \mathbf{S}_{1:t-1}, W^{(1)}, \mathbf{S}[t]=\mathbf{s}  ) - 2NM \log(\pi e) \nonumber\\
& \leq  \sum_{t=1}^{N} \sum_{ \mathbf{s}  }   \mathbb{P}(\mathbf{S}[t]=\mathbf{s})  h(\hat{\mathbf{X}}^{(2)}[t]\oplus \mathbf{Z}^{(2)}[t], \tilde{\mathbf{V}}^{(1)}[t]  \oplus \mathbf{Z}^{(1)}[t] |  \mathbf{Y}^{(1)}_{1:t-1} , \mathbf{S}_{1:t-1} , W^{(1)}  ) - 2NM \log(\pi e) \nonumber\\
& = \sum_{t=1}^{N} h(\hat{\mathbf{X}}^{(2)}[t]\oplus \mathbf{Z}^{(2)}[t], \tilde{\mathbf{V}}^{(1)}[t]  \oplus \mathbf{Z}^{(1)}[t] |  \mathbf{Y}^{(1)}_{1:t-1}  ,\mathbf{S}_{1:t-1}, W^{(1)}  ) - 2NM \log(\pi e) \nonumber\\
& = \sum_{t=1}^{N} h(\hat{\mathbf{X}}^{(2)}[t]\oplus \mathbf{Z}^{(2)}[t], \tilde{\mathbf{V}}^{(1)}[t]  \oplus \mathbf{Z}^{(1)}[t] |  \mathbf{V}^{(1)}_{1:t-1} \oplus \mathbf{Z}^{(1)}_{1:t-1},   \mathbf{S}_{1:t-1}, W^{(1)}  ) - 2NM \log(\pi e) \label{eq:GN_causal_b}
\end{align}
Using inequalities (\ref{eq:GN_causal_a}) and (\ref{eq:GN_causal_b}),
\begin{align}
& N R^{(1)}-  N \epsilon + pN R^{(2)} - pN\epsilon \nonumber \\
& \leq  N  \sum_{j=1}^{M} (1-p) \log \left ( 1+|g_{D,j}|^2   \right )  + p \log   \left ( 1+ \left ( |g_{D,j}| + |g_{I,j}|  \right )^2  \right) + NM \log (\pi e) \nonumber\\ &\quad  - p \sum_{t=1}^{N}  h( \tilde{\mathbf{V}}^{(1)}[t] \oplus  \mathbf{Z}^{(1)}[t] | W^{(1)} ,\mathbf{V}^{(1)}_{1:t-1}\oplus \mathbf{Z}^{(1)}_{1:t-1},  \mathbf{S}_{1:t-1}) - N M (1-p)\log(\pi e) \nonumber \\ & \quad  + p\sum_{t=1}^{N} h(\hat{\mathbf{X}}^{(2)}[t]\oplus \mathbf{Z}^{(2)}[t], \tilde{\mathbf{V}}^{(1)}[t]  \oplus \mathbf{Z}^{(1)}[t] |   \mathbf{V}^{(1)}_{1:t-1} \oplus \mathbf{Z}^{(1)}_{1:t-1}  ,\mathbf{S}_{1:t-1}, W^{(1)}  ) - 2pNM \log(\pi e)\nonumber 
\end{align}
\begin{align}
& = N  \sum_{j=1}^{M} (1-p) \log \left ( 1+|g_{D,j}|^2   \right )  + p \log   \left ( 1+ \left ( |g_{D,j}| + |g_{I,j}|  \right )^2  \right) \nonumber\\ 
&\quad  + p\sum_{t=1}^{N} h(\hat{\mathbf{X}}^{(2)}[t]\oplus \mathbf{Z}^{(2)}[t] | \tilde{\mathbf{V}}^{(1)}[t]  \oplus \mathbf{Z}^{(1)}[t] , \mathbf{V}^{(1)}_{1:t-1} \oplus \mathbf{Z}^{(1)}_{1:t-1}  , \mathbf{S}_{1:t-1}, W^{(1)}  ) - pNM \log(\pi e)\nonumber \\
& \leq N  \sum_{j=1}^{M} (1-p) \log \left ( 1+|g_{D,j}|^2   \right )  + p \log   \left ( 1+ \left ( |g_{D,j}| + |g_{I,j}|  \right )^2  \right) \nonumber\\ 
&\quad  + p\sum_{t=1}^{N} \sum_{j=1}^M h( g_{D,j}x_j^{(2)}[t] + z_j^{(2)}[t] | g_{I,j}x_j^{(2)}[t] + z_j^{(1)}[t]) - pNM \log(\pi e)\nonumber \\
& \leq N  \sum_{j=1}^{M} (1-p) \log \left ( 1+|g_{D,j}|^2   \right )  + p \log   \left ( 1+ \left ( |g_{D,j}| + |g_{I,j}|  \right )^2  \right) \nonumber\\ 
&\quad  + p\sum_{t=1}^{N} \sum_{j=1}^M \log \left ( \pi e \left ( 1+ \frac{|g_{D,j}|^2}{1+ |g_{I,j}|^2} \right) \right ) - pNM \log(\pi e) \nonumber\\
& = N  \sum_{j=1}^{M} (1-p) \log \left ( 1+|g_{D,j}|^2   \right )  + p \log   \left ( 1+ \left ( |g_{D,j}| + |g_{I,j}|  \right )^2  \right)  + p \log \left (  1+ \frac{|g_{D,j}|^2}{1+ |g_{I,j}|^2} \right )  \nonumber\\
& = N   \left(  (1+p) \sum_{j=1}^{M} \log \left ( 1+|g_{D,j}|^2   \right )  +  p  \Delta_{G} \right) 
\end{align}
where $\Delta_G =  \sum_{j=1}^M  \log   \left ( 1+ \left ( |g_{D,j}| + |g_{I,j}|  \right )^2  \right)  +  \log \left (  1+ \frac{|g_{D,j}|^2}{1+ |g_{I,j}|^2} \right ) -2 \log \left ( 1+|g_{D,j}|^2   \right ) $. The bound on $p R^{(1)} + R^{(2)}$ follows by symmetry, and this completes the proof of outer bound (\ref{eq:causal_GN}). \\

\subsubsection{Proof of outer bound (\ref{eq:OB_sum_GN})}
See Appendix~\ref{sec:OB_sum_GN}.
\section{Inner bounds: LD setup} \label{sec:inner_bounds}
In this section, we focus on schemes for achieving the symmetric capacity in the LD setup (see Appendix \ref{sec:D1D2_ach} and \ref{sec:Q1Q2_ach} for achievability of remaining corner points in Figure~\ref{fig:crossover}). In Section~\ref{sec:single_carrier}, we briefly review the single carrier schemes in \cite{ihsiang_bursty_feedback} and describe a \textit{bursty relaying} technique (used in our multicarrier schemes). In Section~\ref{sec:separability}, we mention the cases where treating subcarriers separately is optimal (\emph{i.e.,} simply copying the optimal single carrier scheme \cite{ihsiang_bursty_feedback} on each subcarrier leads to the symmetric capacity). For the remaining cases, we propose multicarrier schemes (covered in Sections~\ref{sec:helping_mechanism} and \ref{sec:insufficient_help}), which employ a helping mechanism where some \textit{helper} levels in subcarriers with $\alpha_j > 2$ are used to recover interfered signals in subcarriers with $\alpha_j < 2$. For $\Delta  \geq 0$ (Section~\ref{sec:helping_mechanism}), the helping mechanism is optimal; whereas for $\Delta < 0$ (Section~\ref{sec:insufficient_help}) the helping mechanism is run in parallel with the single carrier schemes \cite{ihsiang_bursty_feedback} to achieve symmetric capacity. After describing our multicarrier schemes, in Section~\ref{sec:LD_inner_examples} we provide some illustrative examples.
\subsection{Single carrier symmetric capacity \cite{ihsiang_bursty_feedback} and bursty relaying} \label{sec:single_carrier}
The single carrier version of our setup (\emph{i.e.,} $M=1$) was studied in \cite{ihsiang_bursty_feedback}. For notational consistency, we use $j=1$ (subcarrier index) in stating the results from \cite{ihsiang_bursty_feedback}. We simply restate below the schemes in \cite{ihsiang_bursty_feedback} for the regimes $\alpha_1 \leq 1$ and $1 < \alpha_1 \leq 2$; but for the regime $\alpha_1 >2$ we mention a slightly different scheme that makes describing our multicarrier schemes in Sections~\ref{sec:helping_mechanism} and \ref{sec:insufficient_help} more convenient.

\begin{figure}[!ht]
\begin{center}
\includegraphics[scale=1.1]{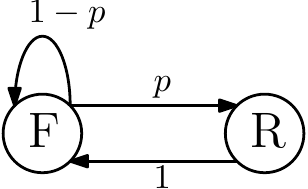}
\caption{Underlying Markov chain for the single carrier schemes in \cite{ihsiang_bursty_feedback} for $\alpha_1 \leq 2$.}
\label{fig:markov_chain}
\end{center}
\end{figure}

\paragraph*{Regime $\alpha_1 \leq 1$}
For this regime, the symmetric capacity is $n_1 -\frac{p}{1+p} k_1 $. To achieve this, a two phase scheme (same for $Tx_1$ and $Tx_2$) is used as briefly described below\footnote{The scheme for $\alpha_1=1$ has slight variation from this scheme. For details, see \cite{ihsiang_bursty_feedback}.} (see \cite{ihsiang_bursty_feedback} for details):
\begin{itemize}
\item Phase $F$: Transmitters in phase $F$ at time index $t$ send fresh symbols on all $n_1$ levels. If there is no interference at time index $t$ (occurs w.p. $1-p$), all $n_1$ symbols can be decoded at the intended receiver and both transmitters stay in phase $F$ for time index $t+1$. If there is interference (occurs w.p. $p$), only the bottom $k_1$ symbols get interfered at a receiver and the transmitters transition to phase $R$ for time index $t+1$.
\item Phase $R$: Transmitters send the past interference (obtained from receiver feedback) on the top $k_1$ levels and fresh symbols on the remaining $(n_1-k_1)$ levels. Both transmitters transition to phase $F$ for the next time index after phase $R$.
\end{itemize}
Figure~\ref{fig:markov_chain} shows the underlying Markov chain for this scheme. 
\paragraph*{Regime $1<\alpha_1 \leq 2$}
For this regime, the symmetric capacity is $\frac{1-p}{1+p}n_1 + \frac{p}{1+p}k_1  = n_1 -\frac{p}{1+p}(2n_1-k_1) $. To achieve this, a two phase scheme is used as briefly described below (see \cite{ihsiang_bursty_feedback} for details):
\begin{itemize}
\item Phase $F$: Transmitters in phase $F$ at time index $t$ send fresh symbols on the top $n_1$ levels and the bottom $k_1-n_1$ levels are not used. If there is no interference at time index $t$ (occurs w.p. $1-p$), all $n_1$ symbols can be decoded at the intended receiver and both transmitters stay in phase $F$. If there is interference at time index $t$ (occurs w.p. $p$), only $2n_1-k_1$ symbols get interfered at a receiver and the transmitters transition to phase $R$ for time index $t+1$.
\item Phase $R$: Transmitters send fresh symbols in the top $k_1-n_1$ levels. In the next $2n_1-k_1$ levels (below the top $k_1-n_1$ levels), the $2n_1-k_1$ interfering symbols (obtained through receiver feedback) from the previous time index are sent. The remaining $k_1-n_1$ levels in the bottom are not used. Both transmitters transition to phase $F$ for the next time index after phase $R$.
\end{itemize}
The underlying markov chain in this scheme is same as the one in Figure~\ref{fig:markov_chain}. 

\begin{figure}[!ht]
\begin{center}
\includegraphics[scale=.5]{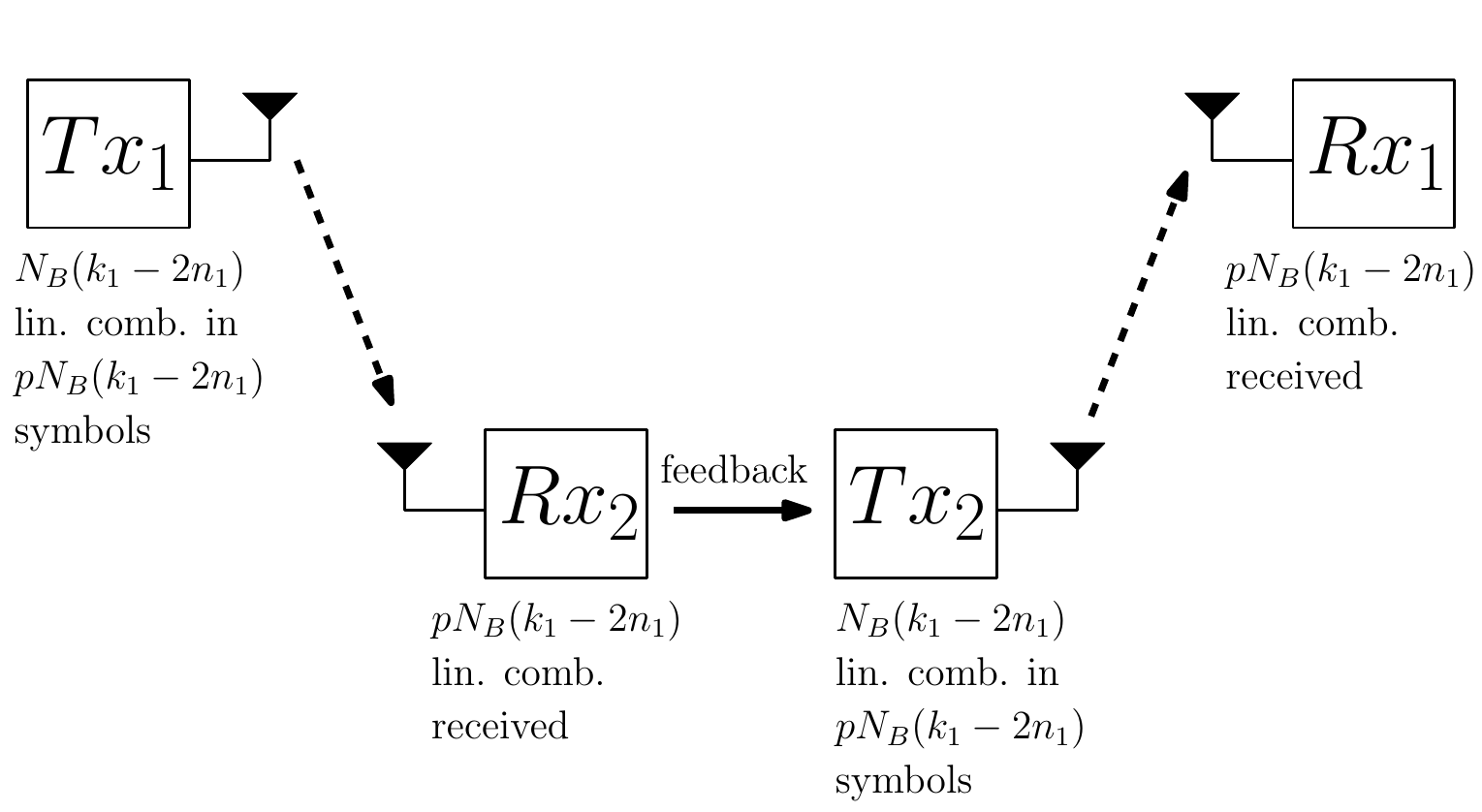}
\caption{Bursty relaying using $k_1-2n_1$ middle levels (below the top $n_1$ levels) when $\alpha_1 >2$. As shown, $Rx_2$ receives $pN_B(k_1-2n_1)$ linear combinations in $pN_B(k_1-2n_1)$ symbols during a block of duration $N_B$. It decodes and sends the constituent symbols to $Tx_2$ which again creates $N_B(k_1-2n_1)$ linear combinations from these symbols. In the next block, $Rx_1$ receives $pN_B(k_1-2n_1)$ linear combinations from $Tx_2$ and decodes the constituent symbols.}
\label{fig:bursty_relaying}
\end{center}
\end{figure}

\paragraph*{Regime $\alpha_1 > 2$ (bursty relaying)}
For this regime, the symmetric capacity is $n_1 + \frac{p}{2}(k_1-2n_1)$. In \cite{ihsiang_bursty_feedback}, this is achieved using a Markov chain based scheme similar to the ones described above. For convenience in describing our multicarrier schemes, we derive a block version of the scheme in \cite{ihsiang_bursty_feedback} as follows. In each block of duration $N_B$, transmitters send fresh symbols on the top $n_1$ levels and never use the bottom $n_1$ levels. Since the bottom $n_1$ levels are never used, the fresh symbols from the top $n_1$ levels are always received interference free. This realizes rate $n_1$. From the $k_1-2n_1$ levels in the middle (below the top $n_1$ levels), we realize an additional rate $\frac{p}{2}(k_1-2n_1)$ over two blocks as follows. For the first block, $Tx_i$ creates $N_B(k_1-2n_1)$ linear combinations from $pN_B(k_1-2n_1)$ fresh symbols and sends these linear combinations in the middle $k_1-2n_1$ levels. For large enough $N_B$, w.h.p. $Rx_{i'}$ receives $pN_B(k_1-2n_1)$ such linear combinations. $Rx_{i'}$ decodes the constituent fresh symbols from these linear combinations and sends them to $Tx_{i'}$ (through feedback). $Tx_{i'}$ now creates $N_B(k_1-2n_1)$ new linear combinations from these symbols and sends them in the $(k_1-2n_1)$ middle levels during the next block. W.h.p. $Rx_{i}$ receives $pN_B(k_1-2n_1)$ such linear combinations and decodes all the constituent symbols. This leads to an additive rate of $\frac{pN_B(k_1-2n_1)}{2N_B} = \frac{p}{2}(k_1-2n_1)$ at $Rx_i$ (and similarly at $Rx_{i'}$). In the remainder of this paper, we refer to this technique (for middle levels in subcarriers with $\alpha_j >2$) as \textit{bursty relaying} since $Tx_i$-$Rx_i$ pair effectively acts as a relay for $Tx_{i'}$-$Rx_{i'}$ and vice versa. Figure~\ref{fig:bursty_relaying} illustrates this technique of bursty relaying. Adding the rate from bursty relaying in $(k_1-2n_1)$ middle levels and rate $n_1$ from the top $n_1$ levels, 
we achieve rate $n_1 + \frac{p}{2}(k_1-2n_1)$.
\subsection{Multicarrier separability} \label{sec:separability}
Using outer bounds (\ref{eq:causal_LD}) and (\ref{eq:OB_3}) for LD setup and achievability rates for the single carrier schemes in \cite{ihsiang_bursty_feedback}, the following can be easily verified. 
\begin{itemize}
\item For $p\in \{0,1\}$ \emph{i.e.,} when interference is either never present or always present, the symmetric capacity can be achieved by treating the subcarriers separately.
\item For $0<p<1$, when all subcarriers have $\alpha_j \leq 2$, the symmetric capacity can be achieved by treating the subcarriers separately.
\item For $0<p<1$, when all subcarriers have $\alpha_j \geq 2$, the symmetric capacity can be achieved by treating the subcarriers separately.
\end{itemize}
Hence, the subcarriers are \textit{separable} in the above cases. When we have subcarriers with $\alpha_j \leq 2$ as well as subcarriers with $\alpha_j >2$ (and $0<p<1$), we employ coding across subcarriers (through a helping mechanism described in the next subsection) to achieve symmetric capacity; we assume such a \textit{mixed} collection of subcarriers in describing our multicarrier schemes in Sections~(\ref{sec:helping_mechanism}) and (\ref{sec:insufficient_help}).
\subsection{Achieving symmetric capacity when $\Delta \geq 0$} \label{sec:helping_mechanism}
When $\Delta \geq 0$, $C_{sym}= R_{NC} = \frac{ p   } {2} \Delta + \sum_{j=1}^{M} n_j$. We will now describe the achievability of $R_{NC}$ using a block based scheme. In each block of duration $N_B$, fresh symbols are sent in the following levels (same for both transmitters by symmetry):
\begin{itemize}
\item All $n_j$ levels for subcarriers with $\alpha_j \leq 1$.
\item Top $n_j$ levels for subcarriers with $\alpha_j > 1$.
\end{itemize}
and the following levels are not used:
\begin{itemize}
\item Bottom $k_j-n_j$ levels of subcarriers with $1<\alpha_j \leq 2 $.
\item Bottom $n_j$ levels of subcarriers with $\alpha_j > 2$.
\end{itemize}
Because of the above choices, in every block (for large enough $N_B$):
\begin{itemize}
\item In subcarriers with $\alpha_j \leq 1$, w.h.p. $pN_B k_j$ fresh symbols get interfered.
\item In subcarriers with $1<\alpha_j \leq 2$, w.h.p. $pN_B (2n_j-k_j)$ fresh symbols get interfered.
\item In subcarriers with $\alpha_j > 2$, the top $n_j$ fresh symbols are always received interference free.
\end{itemize}
In total, each receiver needs to recover $pN_B(\sum_{j:\alpha_j \leq 1}k_j  + \sum_{j: 1 < \alpha_j \leq 2} 2n_j- k_j )$ interfered symbols in each block. This recovery is done in a pipelined fashion in the next block using a \textit{helping mechanism} described below.
\paragraph*{Helping mechanism}
We will use the term \textit{helper levels} for the middle $k_j-2n_j$ levels (below the top $n_j$ levels) in subcarriers with $\alpha_j >2$; hence $\sum_{j:\alpha_j>2}k_j-2n_j$ helper levels in total. After each block, due to feedback from $Rx_i$, $Tx_i$ knows exactly which of its transmitted symbols caused interference at $Rx_{i'}$. The number of such symbols, as described above, is w.h.p. equal to $pN_B(\sum_{j:\alpha_j \leq 1}k_j  + \sum_{j: 1 < \alpha_j \leq 2} 2n_j- k_j )$. $Tx_i$ now creates $N_B(\sum_{j:\alpha_j \leq 1}k_j  + \sum_{j: 1 < \alpha_j \leq 2} 2n_j- k_j )$ linear combinations of these symbols and sends the linear combinations on any $(\sum_{j:\alpha_j \leq 1}k_j  + \sum_{j: 1 < \alpha_j \leq 2} 2n_j- k_j )$ of the helper levels in the subsequent block. W.h.p. $pN_B(\sum_{j:\alpha_j \leq 1}k_j  + \sum_{j: 1 < \alpha_j \leq 2} 2n_j- k_j )$ of such linear combinations are received at $Rx_{i'}$. This is sufficient to recover all the interfered symbols at $Rx_{i'}$ in the previous block.

\par As all the interfered symbols in a block are recovered using the above mechanism, we realize rate $\sum_{j=1}^M n_j$. If $\Delta > 0$, some of the helper levels are still available; precisely $(\sum_{j:\alpha_j > 2}k_j-2n_j) - (\sum_{j:\alpha_j \leq 1}k_j  + \sum_{j: 1 < \alpha_j \leq 2} 2n_j- k_j )=\Delta$ of them. We realize an additional rate of $\frac{p}{2}\Delta$ from such leftover helper levels using the bursty relaying scheme described in Section~\ref{sec:single_carrier}. Adding the rate from the leftover helper levels to $\sum_{j=1}^M n_j$, we achieve the symmetric capacity $\frac{ p   } {2} \Delta + \sum_{j=1}^{M} n_j$.

\subsection{Achieving symmetric capacity when $\Delta < 0$} \label{sec:insufficient_help}
When $\Delta < 0$, $C_{sym}= R_{C} = \frac{ p   } {1+p } \Delta + \sum_{j=1}^{M} n_j $. Before we proceed to the details, we give a high level idea of the scheme as follows. Simply copying the scheme for $\Delta \geq 0$ in Section~\ref{sec:helping_mechanism} does not work for this case since there are not enough helper levels ($\sum_{j:\alpha_j >2}k_j-2n_j$) compared to the number of levels facing interference ($\sum_{j:\alpha_j \leq 1}k_j  + \sum_{j: 1 < \alpha_j \leq 2} 2n_j- k_j $). The trick in this case is to \textit{help as much as possible}. For each subcarrier with $\alpha_j < 2$, we select $h_j$ \textit{helped} levels; these levels face interference and the interfered symbols are recovered using the helping mechanism described in Section~\ref{sec:helping_mechanism}. The total number of helped levels $\sum_{j:\alpha_j <2} h_j$ equals the number of helper levels ($\sum_{j:\alpha_j >2} k_j-2n_j$).
For the remaining interfered levels in subcarriers with $\alpha_j <2$, we run the optimal single carrier scheme \cite{ihsiang_bursty_feedback} (with a slight modification) in parallel with the helping mechanism. Adding the rates from the single carrier schemes and the helping mechanism, we achieve the symmetric capacity. This high level idea can also be illustrated by rewriting $R_C = \frac{ p   } {1+p } \Delta + \sum_{j=1}^{M} n_j$ as shown below.
\begin{align}
& \frac{p}{1+p}\Delta + \sum_{j=1}^{M} n_j  \nonumber\\
&= \left (\sum_{j:\alpha_j \geq 2} n_j \right )  + \left ( \sum_{j:\alpha_j < 2} h_j \right ) \nonumber +  \left (\sum_{j:\alpha_j \leq 1} (n_j- h_j) - \frac{p}{1+p}(k_j-h_j) \right ) +  \left ( \sum_{j:1<\alpha_j < 2} (n_j- h_j) - \frac{p}{1+p}(2(n_j-h_j)-(k_j-h_j) ) \right )  \nonumber\\
&= \left (\sum_{j:\alpha_j \geq 2} n_j \right )  + \left ( \sum_{j:\alpha_j < 2} h_j \right )  +  \left ( \sum_{j:\alpha_j \leq 1} \tilde{n}_j - \frac{p}{1+p}\tilde{k}_j  \right )   +   \left (\sum_{j:1<\alpha_j < 2} \tilde{n}_j - \frac{p}{1+p}(2\tilde{n}_j-\tilde{k}_j ) \right ) \label{eq:rate_split}
\end{align}
where $\sum_{j:\alpha_j < 2} h_j = \sum_{j:\alpha_j >2} k_j-2n_j $ is the total number of helped levels, and for subcarriers (with $\alpha_j <2$) being helped the effective direct and interfering link strengths are $\tilde{n}_j = n_j-h_j$ and $\tilde{k}_j = k_j-h_j$. The last two terms in (\ref{eq:rate_split}) come from the optimal single carrier schemes for $\alpha_j <2$ (that run in parallel with the helping mechanism).

\par We now describe the achievability of $R_{C}$ in detail. In subcarriers with $\alpha_j \geq 2$, the transmitters always send fresh symbols in the top $n_j$ levels and never use the bottom $n_j$ levels. This realizes rate $\sum_{j:\alpha_j \geq 2} n_j$. For each subcarrier with $\alpha_j < 2$, we assign a non-negative integral value $h_j$ with the following constraints: ($a$) $h_j \leq k_j$ for $\alpha _j \leq 1$ and  $h_j \leq 2n_j-k_j$ for $1<\alpha _j < 2$, ($b$) $\sum_{j:\alpha_j < 2}h_j = \sum_{j:\alpha_j > 2} k_j -2n_j$. Simply put, $h_j$ denotes the number of helped levels in a subcarrier and the total number of such levels equals the number of helper levels available in subcarriers with $\alpha_j >2$. Having fixed $h_j$ for each subcarrier with $\alpha_j <2$, we now describe the modifications needed in the optimal single carrier scheme \cite{ihsiang_bursty_feedback} for parallel execution with the helping mechanism.

\begin{figure*}[!ht]
\begin{center}
\includegraphics[scale=.85]{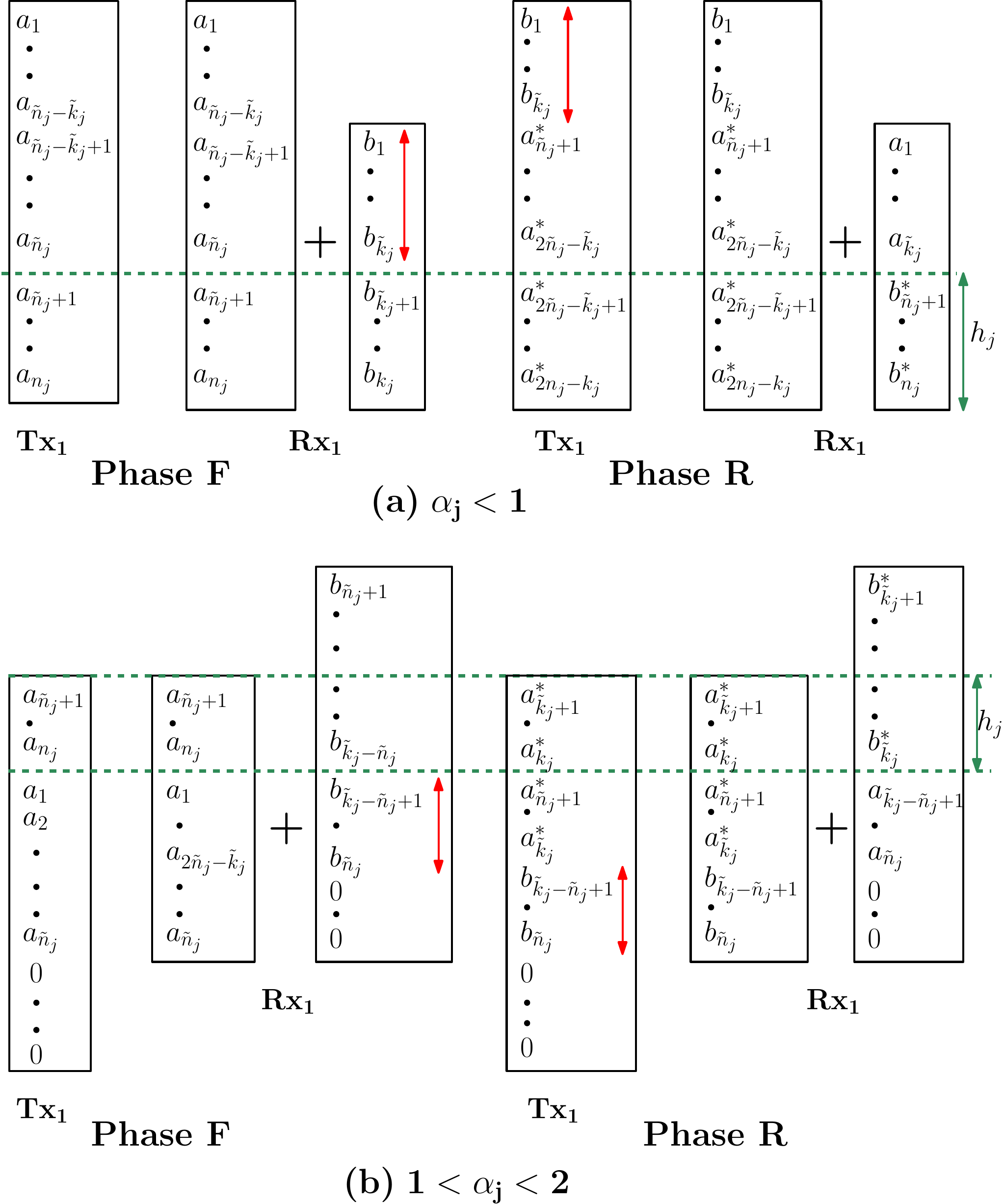}
\caption{Modified single carrier schemes for $\alpha_j <1$ and $1<\alpha_j <2$ which run in parallel with the helping mechanism when $\Delta <0$. Because of $h_j$ helped levels, the effective direct and interfering link strengths are $\tilde{n}_j = n_j - h_j $ and $\tilde{k}_{j}= k_j-h_j$. The bidirectional red arrows indicate the interfering symbols (from phase $F$) sent in phase $R$ of the modified scheme.}
\label{fig:parallel}
\end{center}
\end{figure*}

\paragraph*{Modification for $\alpha_j < 1$} The bottom $h_j$ levels (of the direct link) are selected as helped levels as shown in Figure~\ref{fig:parallel}(a) and interfered symbols in these levels are recovered using the helping mechanism described in Section~\ref{sec:helping_mechanism}. For the modified single carrier scheme, phase $F$ remains the same as in \cite{ihsiang_bursty_feedback} and the modification is only in Phase $R$. For illustration purposes consider that in phase $F$ for a subcarrier with $\alpha_j <1$, $Tx_1$ sends fresh symbols $[a_1 \; a_2\; \ldots a_{n_j}]$ (as shown in Figure~\ref{fig:parallel}(a)) and $Tx_2$ sends fresh symbols $[b_1 \; b_2\; \ldots b_{n_j}]$. If there is no interference, all the fresh symbols are received and the transmitters stay in phase $F$. If there is interference, the transmitters transition to phase $R$. In the scheme in \cite{ihsiang_bursty_feedback}, all $k_j$ interfering symbols were sent on the top $k_j$ levels in phase $R$; in the modified scheme the transmitters just send the top $\tilde{k}_{j}=k_j-h_j$ interfering symbols in the top $\tilde{k}_j$ levels as shown in Figure~\ref{fig:parallel}(a). In the remaining levels, fresh symbols are sent (starred symbols in Figure~\ref{fig:parallel}(a)). Ignoring the bottom $h_j$ levels, the resulting system of linear equations at the receivers is exactly the same as in \cite{ihsiang_bursty_feedback} with direct link strength $\tilde{n}_j$ and interfering link strength $\tilde{k}_{j}$. Thus at end of phase $R$, $Rx_1$ is able to decode $\{a_{\tilde{n}_j-\tilde{k}_j +1 }, a_{\tilde{n}_j-\tilde{k}_j +2} , \ldots a_{\tilde{n}_j}  \}$ (interfered symbols in phase $F$) and $\{a^*_{\tilde{n}_j +1 }, a^*_{\tilde{n}_j +2} , \ldots a^*_{2\tilde{n}_j-\tilde{k}_j} \}$ (fresh symbols in phase $R$). To decode interfered symbols in the helped  levels, the helping mechanism is used (which collects all interfered symbols in helped levels during a block of duration $N_B$ and enables their recovery in the next block). So effectively, the rate obtained from a subcarrier with $\alpha_j \leq 1$ is $h_j + (\tilde{n}_j - \frac{p}{1+p}\tilde{k}_j)$.

\paragraph*{Modification for $\alpha_j = 1$} The case $k_j=n_j$ is just an aggregated version of the simple case $k_j=n_j=1$. For this simple case, either $h_j=0$ or $h_j=1$. If $h_j=1$, we use the helping mechanism to recover the interfered symbols. If $h_j=0$, there are no helped levels and we simply use the scheme for $\alpha_j=1$ in \cite{ihsiang_bursty_feedback}.

\paragraph*{Modification for $1<\alpha_j <2$} The top $h_j$ levels (of the direct link at the receiver) are selected as helped levels as shown in Figure~\ref{fig:parallel}(b). Again, phase $F$ remains the same as in \cite{ihsiang_bursty_feedback} and the modification is only for phase $R$. For illustration purposes, consider that in phase $F$ for a subcarrier with $1<\alpha_j <2$, $Tx_1$ sends fresh symbols $[a_{\tilde{n}_j+1} \; a_{\tilde{n}_j+2}\; \ldots a_{n_j}\;a_1 \;a_2\ldots a_{\tilde{n}_j}  ]$ on the top $n_j$ levels\footnote{This particular labeling of the symbols is just for convenience in describing the modification in phase $R$.} (as shown in Figure~\ref{fig:parallel}(b)). Similarly, $Tx_2$ sends fresh symbols $[b_{\tilde{n}_j+1} \;b_{\tilde{n}_j+2}\; \ldots b_{n_j}\;b_1 \;b_2\ldots b_{\tilde{n}_j}  ]$ on the top $n_j$ levels. The bottom $k_j-n_j$ levels are not used. If there is no interference, all the fresh symbols are received and the transmitters stay in phase $F$. If there is interference, the transmitters transition to phase $R$. In phase $R$ of the scheme in \cite{ihsiang_bursty_feedback}, the bottom $k_j-n_j$ levels were not used and the $2n_j-k_j$ interfering symbols in phase $F$ were sent on the $2n_j-k_j$ levels above the unused levels. In the modified scheme, the transmitters send only $2n_j-k_j-h_j = 2\tilde{n}_j-\tilde{k}_j$ interfering symbols (from phase $F$) on the $2\tilde{n}_j-\tilde{k}_j$ levels above the $k_j-n_j$ unused levels in the bottom. These interfering symbols correspond to the $2\tilde{n}_j-\tilde{k}_j$ levels below the top $h_j$ levels in the direct link at the receiver as shown in Figure~\ref{fig:parallel}(b). In the remaining levels, fresh symbols are sent (starred symbols in Figure~\ref{fig:parallel}(b)). Ignoring the $h_j$ helped levels, the resulting system of linear equations at the receivers is exactly the same as in \cite{ihsiang_bursty_feedback} with direct link strength $\tilde{n}_j$ and interfering link strength $\tilde{k}_{j}$. Thus at end of phase $R$, $Rx_1$ is able to decode $\{a_{1}, a_{2} , \ldots a_{2\tilde{n}_j-\tilde{k}_j}  \}$ (interfered symbols in phase $F$) and $\{a^*_{\tilde{n}_j +1 }, a^*_{\tilde{n}_j +2} , \ldots a^*_{\tilde{k}_j} \}$ (fresh symbols in phase $R$). To decode interfered symbols in the helped  levels, the helping mechanism is used (which collects all interfered symbols in helped levels during a block of duration $N_B$ and enables their recovery in the next block). So effectively, the rate obtained from a subcarrier with $1<\alpha_j < 2$ is $h_j + (\tilde{n}_j - \frac{p}{1+p}(2\tilde{n}_j-\tilde{k}_j))$.
\par Taking into account the above modifications and adding the rates across subcarriers we achieve rate $R_C$.
\subsection{Toy example revisited} \label{sec:LD_inner_examples}
\begin{figure}[htp]
\centering
\includegraphics[scale=0.54]{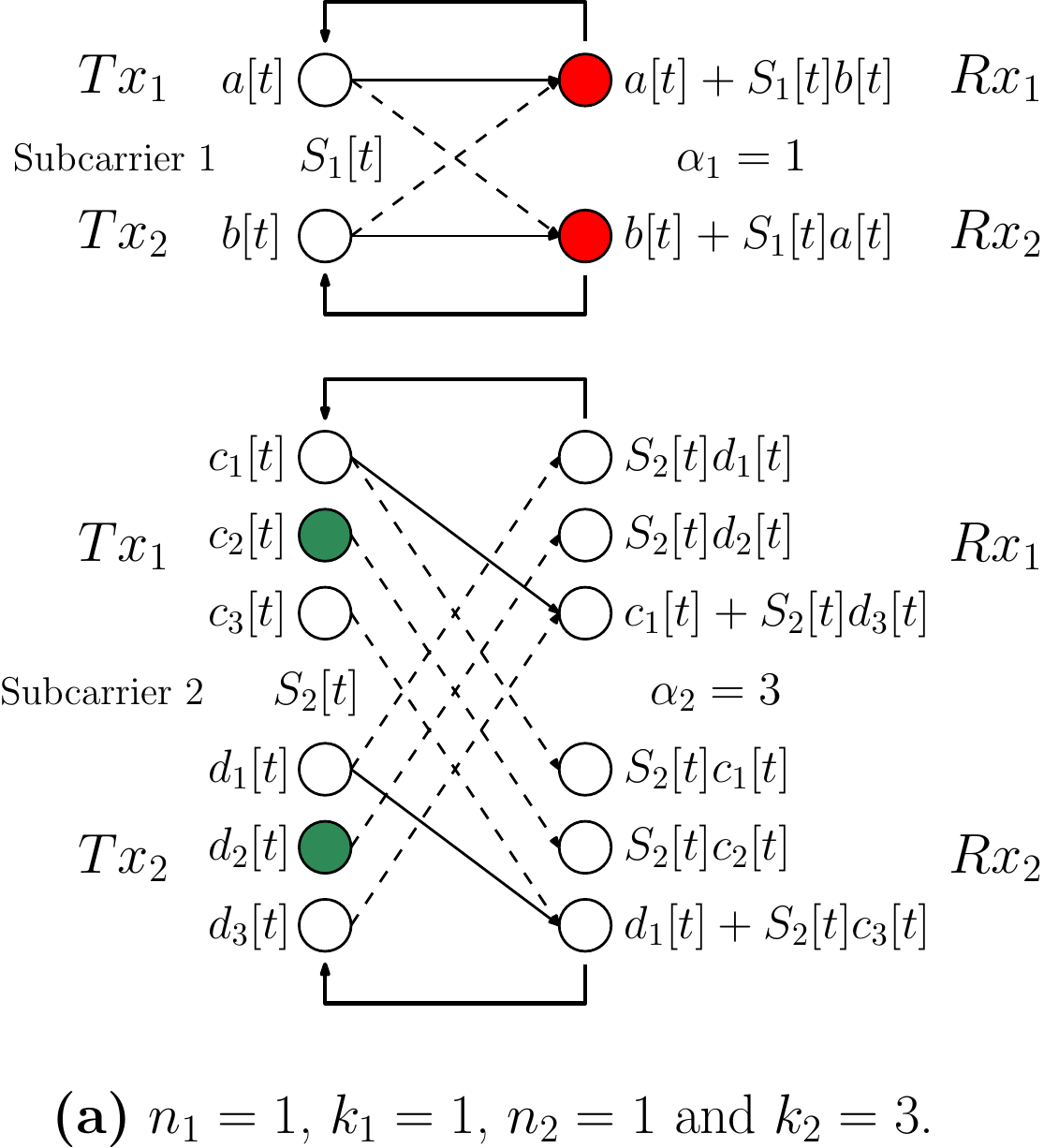}\hfill
\includegraphics[scale=0.54]{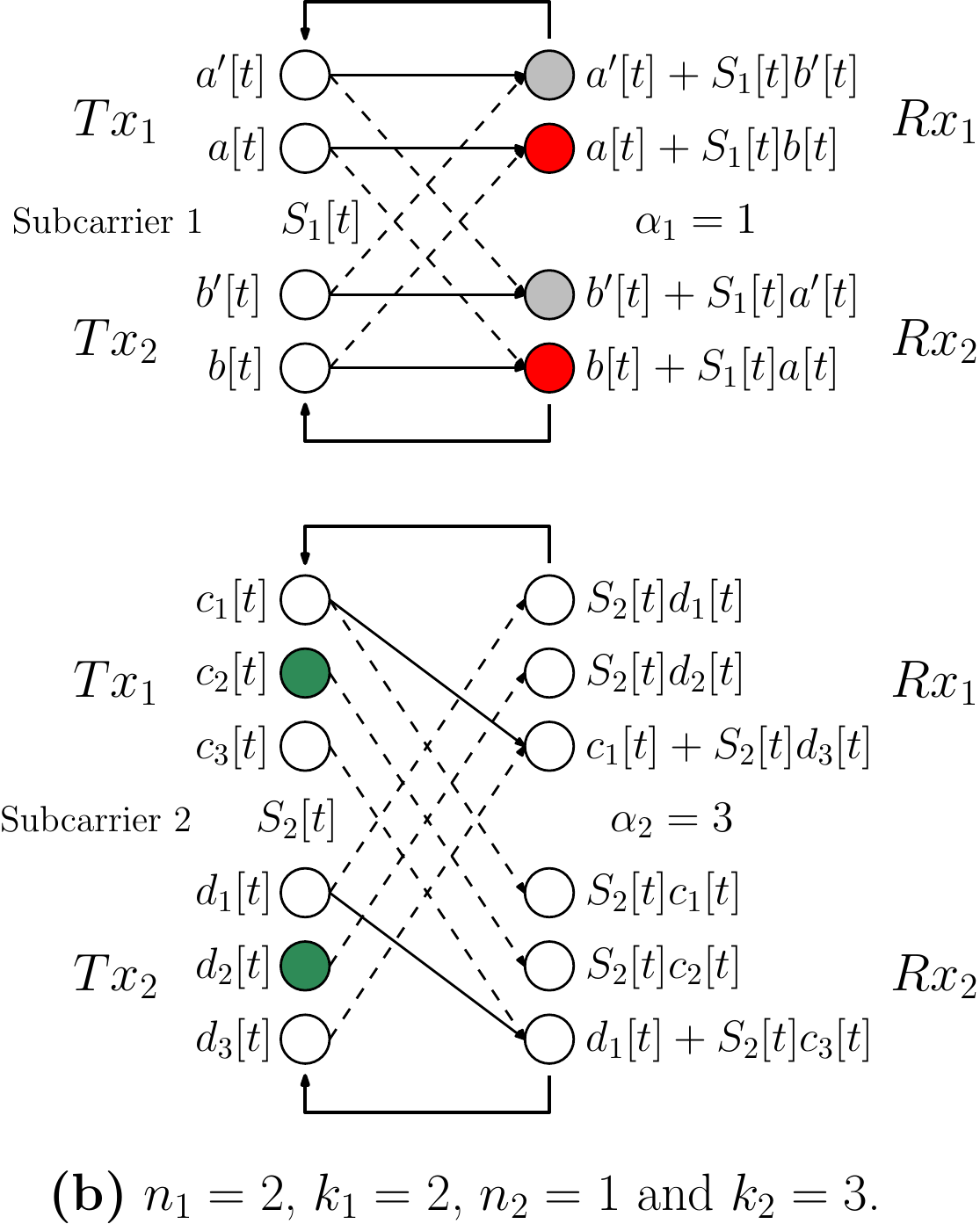}\hfill
\includegraphics[scale=0.54]{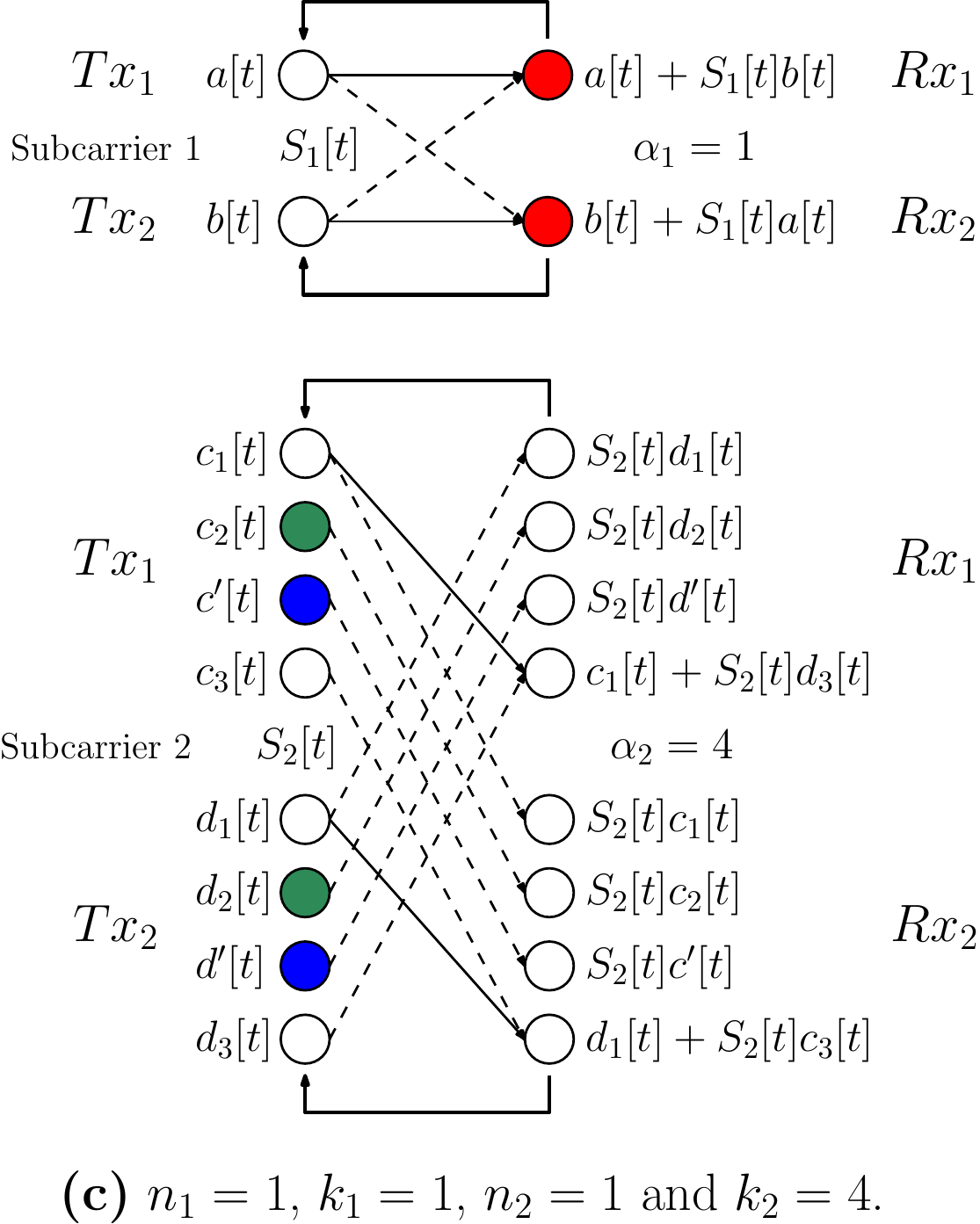}
\caption{Toy example and its modifications: (a) original toy example, (b) Example~\ref{ex:delta_neg} and (c) Example~\ref{ex:delta_pos}. }
\label{fig:toy_example_revisited}
\end{figure}

The toy example in Section~\ref{sec:introduction} considered two subcarriers with $n_1=1$, $k_1=1$, $n_2=1$ and $k_2=3$ (and $p=\frac{1}{2}$). As illustrated in the toy example, the middle level in the second subcarrier helped in recovering interfered symbols in the first subcarrier.
With reference to our achievability scheme for $\Delta \geq 0$, the middle level in the second subcarrier is a helper level (green level in Figure~\ref{fig:toy_example_revisited}(a)) whereas the (only) level in the first subcarrier is a helped level (red level in Figure~\ref{fig:toy_example_revisited}(a)).
Since there is only one helped level and one helper level, $\Delta=1-1=0$ and $C_{sym}=2$.
To illustrate ideas behind our achievability schemes for $\Delta >0$ and $\Delta <0$, we slightly modify the toy example as described below.
\begin{example}[$n_1=2$, $k_1=2$, $n_2=1$ and $k_2=3$] \label{ex:delta_neg}
Compared to the original toy example, we have modified only the first subcarrier. For this case, there are two levels in the first subcarrier which face may interference but there is only one helper level (green level in Figure~\ref{fig:toy_example_revisited}(b)) available in the second subcarrier. Hence $\Delta= 1-2 =-1$ and $C_{sym}=2+\frac{2}{3}$. We help the bottom level in the first subcarrier (as we did in the original toy example) and by simply copying the scheme in the original toy example we achieve rate $2$. For the top level in the first subcarrier (gray level in Figure~\ref{fig:toy_example_revisited}(b)), we use the optimal single carrier scheme for $\alpha_1=1$ \cite{ihsiang_bursty_feedback} and achieve additional rate $\frac{2}{3}$. In this example, it is easy to see that the helping mechanism and the single carrier scheme can be executed in parallel.
\end{example}

\begin{example}[$n_1=1$, $k_1=1$, $n_2=1$ and $k_2=4$] \label{ex:delta_pos}
Compared to the original toy example, we have modified only the second subcarrier such that it has one extra middle level (blue level in Figure~\ref{fig:toy_example_revisited}(c)). For this case, $\Delta = 2-1 =1$ and $C_{sym}= 2+ \frac{1}{4}$. The helping mechanism is used as in the original toy example to achieve rate $2$. Additional rate $\frac{1}{4}$ is achieved using the bursty relaying technique for the extra middle level in the second subcarrier (blue level in Figure~\ref{fig:toy_example_revisited}(c)).
\end{example}
\section{GDoF: GN setup} \label{sec:gdof}
In this section, we first describe tight outer bounds (described below) followed by tight inner bounds (in Sections~\ref{sec:GDoF_inner_pos} and \ref{sec:GDoF_inner_neg}) on the GDoF for GN setup. As mentioned in Section~\ref{sec:main_results}, for the GDoF analysis we assume $g_{D,j} = \sqrt{SNR}$,  $g_{I,j} = \sqrt{INR_j}$ and $INR_j = SNR^{\beta_j}$. We assume a rational $\beta_j$ to simplify the achievability schemes (described in Sections~\ref{sec:GDoF_inner_pos} and \ref{sec:GDoF_inner_neg}). With the above assumptions, the GDoF for GN setup is defined as follows,
\begin{align}
GDoF\left( \beta_1, \beta_2,\ldots \beta_M \right) & = \limsup_{SNR \rightarrow \infty} \frac{ C_{sym} \left( SNR, \beta_1,\beta_2, \ldots \beta_M \right) }{M \log(SNR)} \nonumber
\end{align}
where $C_{sym} $ is the symmetric capacity.
From outer bounds (\ref{eq:causal_GN}) and (\ref{eq:OB_sum_GN}) for the GN setup, we have bounds on $C_{sym}$ as follows,
\begin{align}
C_{sym} & \leq \min \left( \frac{p}{2} \Delta_{G} + \sum_{j=1}^{M} \log \left ( 1+|g_{D,j}|^2   \right )     , \; \;  \frac{p}{1+p} \Delta_{G}  + \sum_{j=1}^{M} \log \left ( 1+|g_{D,j}|^2   \right )   \right) \nonumber\\
&=\begin{cases}
    \frac{p}{2} \Delta_{G} + \sum_{j=1}^{M} \log \left ( 1+|g_{D,j}|^2 \right) & \text{if $ \Delta_G \geq 0$}.\\
    \frac{p}{1+p} \Delta_{G} + \sum_{j=1}^{M} \log \left ( 1+|g_{D,j}|^2\right)  & \text{if $ \Delta_G < 0$}. \label{eq:C_sym_bound}
  \end{cases}
\end{align}
where $\Delta_G =  \sum_{j=1}^M  \log   \left ( 1+ \left ( |g_{D,j}| + |g_{I,j}|  \right )^2  \right)  +  \log \left (  1+ \frac{|g_{D,j}|^2}{1+ |g_{I,j}|^2} \right ) -2 \log \left ( 1+|g_{D,j}|^2   \right )$. Using (\ref{eq:C_sym_bound}), the following outer bound on GDoF holds,
\begin{align}
GDoF(\beta_1,\ldots, \beta_M) &\leq \min \left( \lim_{SNR \rightarrow \infty} \frac{ \frac{p}{2} \Delta_{G} + \sum_{j=1}^{M} \log \left ( 1+|g_{D,j}|^2 \right)} {M \log (SNR) }, \; \lim_{SNR \rightarrow \infty} \frac{ \frac{p}{1+p} \Delta_{G} + \sum_{j=1}^{M} \log \left ( 1+|g_{D,j}|^2 \right)} {M \log (SNR) }  \right)\nonumber\\
&= \min \left( \frac{\frac{p}{2} \Delta_{GDoF}}{M} +1  ,  \frac{\frac{p}{1+p} \Delta_{GDoF}}{M} +1  \right) \label{eq:GDoF_OB}
\end{align}
where $\Delta_{GDoF} = \lim_{SNR \rightarrow \infty} \frac{\Delta_G}{\log \left( SNR \right)} = \sum_{j=1}^M  \left( \max \left( 1, \beta_j \right)  +  (1-\beta_j)^{+}  - 2  \right) = \left(  \sum_{j:\beta_j > 2} \beta_j - 2 \right) - \left (\sum_{j:\beta_j \leq 1 } \beta_j \right) - \left(  \sum_{j:1 < \beta_j \leq 2} 2- \beta_j  \right)$. \\
\par In the remainder of this section, we describe achievability schemes (inner bounds) which achieve outer bound (\ref{eq:GDoF_OB}).
The schemes for the GDoF setting mimic the achievability schemes for symmetric capacity in the LD setup by using techniques from \cite{jafar_gdof}.
Hence, the scheme for $\Delta_{GDoF} \geq 0$ (Section~\ref{sec:GDoF_inner_pos}) in the GDoF setting mimics the scheme for $\Delta \geq 0$ in LD setup and the scheme for $\Delta_{GDoF} < 0$ (Section~\ref{sec:GDoF_inner_neg}) mimics the scheme for $\Delta < 0$ in LD setup.
\subsection{GDoF inner bound when $\Delta_{GDoF} \geq 0$ } \label{sec:GDoF_inner_pos}
We use a block based scheme (block size $N_B$) which mimics the scheme for $\Delta \geq 0$ in Section~\ref{sec:helping_mechanism} for LD setup.
For convenience in describing our scheme, we will work with the following \textit{real} channel (the achievable rate for the complex channel in GN setup is just twice the achievable rate for this channel).
\begin{align}
y^{(i)}_j[t] &= \sqrt{SNR}\; x^{(i)}_j[t] + \left( S_j[t] \right) \sqrt{INR_j}\; x^{(i')}_j[t]    + z^{(i)}_j[t]  \label{eq:GDoF_real_channel}
\end{align}
where $ x^{(i)}_j[t],\;  x^{(i')}_j[t] \in \mathbb{R}$, $\frac{1}{N} \sum_{t=1}^N |x^{(i)}_j[t]|^2 \leq 1 $ and $ z^{(i)}_j[t] \sim \mathcal{N}(0,1)$. Similar to the analysis in \cite{jafar_gdof}, we consider
\begin{align}
SNR = Q^{2m} \label{eq:GDoF_SNR}
\end{align}
where $Q$ and $m$ are positive integers. Furthermore, $m$ is such that $\forall j\in\{1,2,\ldots M\}, \; m\beta_j$ is an integer (always possible since all $\beta_j$ are rational). By letting $m$ grow to infinity, we get a sequence of SNRs that approach infinity. Using (\ref{eq:GDoF_SNR}), the received signal in (\ref{eq:GDoF_real_channel}) can be rewritten as follows.
\begin{align}
y^{(i)}_j[t] &= Q^m x^{(i)}_j[t] + \left( S_j[t] \right) Q^{m\beta_j} x^{(i')}_j[t]    + z^{(i)}_j[t] \label{eq:GDoF_Q_channel}
\end{align}
Following \cite{jafar_gdof}, we will express positive real signals in $Q$-ary representation using $Q$-ary digits $0,1,\ldots Q-1$ (which we will refer to as ``qits'', similar to \cite{jafar_gdof}).
To mimic the achievability scheme for $\Delta \geq 0$ in LD setup (Section~\ref{sec:helping_mechanism}), we use the following structure for the input signals (we drop the time index for convenience).
\begin{itemize}
\item For $j:\beta_j >2$, 
\begin {align} 
x^{(i)}_j = \left [0\;. \; x^{(i)}_{j,m\beta_j} \; x^{(i)}_{j,m\beta_j -1 } \; \ldots x^{(i)}_{j,1}  \right]_Q
\end{align}
where $x^{(i)}_{j,1}  = x^{(i)}_{j,2} = \ldots = x^{(i)}_{j,m} = 0$ and for the remaining $r \in \{1,2,\ldots m\beta_j\} -\{1,2,\ldots m\}$, $x^{(i)}_{j,r} \in  \{1,2,\ldots Q-2\}$.  \label{eq:struct_2}
\item For $j: \beta_j \leq 1 $, 
\begin{align} 
x^{(i)}_j = \left [0\;. \; x^{(i)}_{j,m} \; x^{(i)}_{j,m-1 } \; \ldots x^{(i)}_{j,1}  \right]_Q  \label{eq:struct_1}
\end{align}
where $ x^{(i)}_{j,r }\in \{1,2,\ldots \lfloor \frac{Q-1}{2} \rfloor -1 \}$ for $r \in \{1,2,\ldots m\}$.
\item For $j: 1 <\beta_j \leq 2 $, 
\begin{align}
x^{(i)}_j = \left [0\;.\; x^{(i)}_{j,m\beta_j} \; x^{(i)}_{j,m\beta_j -1 } \; \ldots x^{(i)}_{j,1}  \right]_Q  \label{eq:struct_1_2}
\end{align}
where $x^{(i)}_{j,1}  = x^{(i)}_{j,2} = \ldots = x^{(i)}_{j,m(\beta_j-1)} = 0$ and for the remaining $r \in \{1,2,\ldots m \beta_j\} -\{1,2,\ldots m (\beta_j-1)\}$, \\ $ x^{(i)}_{j,r} \in \{1,2,\ldots \lfloor \frac{Q-1}{2} \rfloor -1 \}$.
\end{itemize}
The structure (\emph{i.e.,} non-zero qits) used is same as in the scheme for LD setup (Section~\ref{sec:helping_mechanism}). The restrictions on the values taken by non-zero qits arises from techniques in \cite{jafar_gdof} (these simplify the analysis by preventing carry overs when signals interfere, see \cite{jafar_gdof} for details). In the absence of noise, it is easy to see the similarities between the LD setup and above setup; qits in a signals are similar to \textit{levels} in the LD setup. The following example makes this similarity more precise for the case of subcarriers with $\beta_j >2$.
\begin{example}
In a subcarrier with $\beta_j >2$, the received signal at $Rx_i$ after interference (in the absence of noise) is as follows.
\begin{align}
\left [x^{(i)}_{j,m\beta_j} \; x^{(i)}_{j,m\beta_j-1} \; \ldots x^{(i)}_{j,m\beta_j-m+1}\; . \; x^{(i)}_{j,m\beta_j-m} \ldots x^{(i)}_{j, 1}    \right]_Q  + \left [x^{(i')}_{j,m\beta_j} \; x^{(i')}_{j,m\beta_j-1} \; \ldots \; x^{(i')}_{j, m+1 } \; 0\;0\;\ldots 0\;.\;0 \;0   \right]_Q 
\end{align}
Clearly, the top $m$ qits of the direct signal (\emph{i.e.,} $x^{(i)}_{j,m\beta_j} \; x^{(i)}_{j,m\beta_j-1} \; \ldots x^{(i)}_{j,m\beta_j-m+1}$) are interference free in the above scenario and by doing a modulo $Q^{m}$ operation at the receiver, one can completely recover the direct signal. Even in the presence of noise, due to bounded variance of the noise, the higher qits can be decoded with negligible probability of error (as $m \rightarrow \infty$).
\end{example}
\par Having shown the similarity between LD setup and the above setup in the absence of noise, we now describe the rates that we can achieve from the subcarriers in the GDoF setting.

\paragraph{$\beta_j \leq 1$}
In this case, over a block only $(p N_B) m \beta_j$ qits in the direct signal are interfered. Assuming we are able to recover all (except $o(m)$) interfering qits (using the helping mechanism described for $\beta_j >2$ below), we can achieve the following rate:
 \begin{align}
m \log_Q \left ( \lfloor \frac{Q-1}{2}\rfloor -1 \right)  + o(m) \nonumber
\end{align}
The above rate follows directly from the analysis in \cite{jafar_gdof}.
\paragraph{$1<\beta_j \leq 2$}
In this case, over a block only $(p N_B) m(2- \beta_j)$ qits in the direct signal are interfered. Assuming we are able to recover all (except $o(m)$) interfering qits (using the helping mechanism described for $\beta_j >2$ below), we can achieve the following rate:
\begin{align}
m \log_Q \left ( \lfloor \frac{Q-1}{2}\rfloor -1 \right)  + o(m) \nonumber
\end{align}
\paragraph{$\beta_j > 2$}
The top $m$ qits in the subcarriers with $\beta_j >2$ are always received interference free. So from them we can achieve rate:
\begin{align}
m \log_Q \left ( Q -2 \right)  + o(m) \nonumber
\end{align}
We now describe the helping mechanism for the GDoF setting. For removing the interfering qits for subcarriers with $\beta_j < 2$ in the previous block, we need to use $\sum_{j:1 <\beta_j \leq 2} m(2-\beta_j) + \sum_{j:\beta_j \leq 1} m\beta_j$ \textit{helper} qits in subcarriers with $\beta_j>2$; these are the middle $m(\beta_j-2)$ qits below the top $m$ qits. Since $\Delta_{GDoF} \geq 0$, we have sufficient number of such helper qits to recover all interfering qits in subcarriers with $\beta_j < 2$. The helping mechanism is same as described for the LD setup (with minor changes for the $Q$-ary setup). From the leftover helper qits, we can achieve an additional rate using the bursty relaying technique. Summing the rates for all subcarriers we have the following inner bound (a factor of $\frac{1}{2}$ is included to account for the complex channel).
\begin{align}
\frac{1}{2}C_{sym}(SNR,\beta_1,\ldots \beta_M)& \geq   \left (m \sum_{j:\beta_j \leq 2}\log_Q \left ( \lfloor \frac{Q-1}{2}\rfloor -1 \right)  + o(m)  \right)  + \left ( m \sum_{j:\beta_j >2 } \log_Q \left ( Q-2\right)  + o(m) \right) \nonumber\\ & \quad + \left( \frac{p}{2} m \left(  \sum_{j:\beta_j >2} (\beta_j-2)  - \sum_{j:\beta_j \leq 1 }\beta_j - \sum_{j:1<\beta_j \leq 2 } (2-\beta_j)  \right) \log_Q \left( Q - 2 \right) + o(m) \right) 
\end{align} 
So, 
\begin{align}
GDoF(\beta_1,\ldots \beta_M )& =  \limsup_{m\rightarrow \infty} \frac{C_{sym}(SNR,\beta_1,\ldots \beta_M)}{M\log_Q\left( Q^{2m}\right)} \nonumber\\
& \stackrel{(a)}\geq \frac{ \frac{p}{2} \left ( \left(  \sum_{j:\beta_j > 2} \beta_j - 2 \right) - \left (\sum_{j:\beta_j \leq 1 } \beta_j \right) - \left(  \sum_{j:1 < \beta_j \leq 2} 2- \beta_j  \right) \right)   }{M} +1  \nonumber\\ 
&=  \frac{ \frac{p}{2} \Delta_{GDoF} }{M} +1  \nonumber
\end{align}
where (a) follows from large enough $Q$. Since the inner bound on GDoF matches the outer bound, we have a tight result when $\Delta_{GDoF}  \geq 0$.
\subsection{GDoF inner bound when $\Delta_{GDoF} < 0$} \label{sec:GDoF_inner_neg}
As in the case of $\Delta_{GDoF} \geq 0$ in Section~\ref{sec:GDoF_inner_pos}, we focus on the real channel in (\ref{eq:GDoF_Q_channel}) for our achievability scheme. The scheme for this case mimics the achievability of symmetric capacity in LD setup for $\Delta < 0$ by using the techniques from \cite{jafar_gdof}. Since we have already illustrated the usage of techniques from \cite{jafar_gdof} (for the case $\Delta_{GDoF} \geq 0$) in mimicking the LD setup schemes for the GDoF setting, we will briefly sketch the inner bound for $\Delta_{GDoF} < 0$.  

\par Following the strategy of \textit{helping as much possible} for the case $\Delta < 0$ in LD setup, we use the middle $m(\beta_j-2)$ qits (below the top $m$ qits) in subcarriers with $\beta_j >2$ as helper qits. All the helper qits are used to recover interference in helped qits in subcarriers with $\beta_j <2$ (each subcarrier with $\beta_j <2$ has $h_j$ helped qits and $\sum_{j:\beta_j <2} h_j = \sum_{j:\beta_j>2}m(\beta_j-2)$). So we get the following rates from subcarriers:
\begin{itemize}
\item For $j:\beta_j \geq 2$ $\rightarrow$ $m \log_Q \left( Q - 2 \right) + o(m)$
\item For $j:1<\beta_j < 2$ $\rightarrow$  $\left( h_j + \frac{1-p}{1+p} (m-h_j) + \frac{p}{1+p} (m\beta_j-h_j) \right)\log_Q \left ( \lfloor \frac{Q-1}{2}\rfloor -1 \right)  + o(m)    $
\item For $j: \beta_j \leq 1$ $\rightarrow$ $\left( h_j + (m -h_j) -\frac{p}{1+p} (m\beta_j-h_j) \right)\log_Q \left ( \lfloor \frac{Q-1}{2}\rfloor -1 \right)  + o(m)  $
\end{itemize}
It should be noted that due to noise, some of the interfering qits in phase $F$ (of the single carrier scheme executed in parallel with the helping mechanism) may not be decoded correctly at $Tx_i$ (after feedback) and this may affect the recovery of qits in phase $R$. However, it can be shown that such an \textit{error propagation} leads to $o(m)$ reduction (compared to the case without noise) in the achievable rate for a subcarrier.
Combining the rates from all subcarriers, we have the following bound (factor of $2$ included for the complex channel).
\begin{align}
C_{sym}(SNR,\beta_1,\ldots \beta_M)& \geq 2  \left( \sum_{j:\beta_j < 2} h_j  +  \sum_{j:\beta_j \geq 2}m + \sum_{j:\beta_j \leq 1} (m-h_j) -\frac{p}{1+p} (m\beta_j-h_j) + \sum_{j:1 < \beta_j < 2}    \frac{1-p}{1+p} (m-h_j) + \frac{p}{1+p} (m\beta_j-h_j) \right) \times \nonumber\\ & \quad \log_Q \left ( \lfloor \frac{Q-1}{2}\rfloor -1 \right)   + o(m) \nonumber\\
&\stackrel{(a)}=2\left( \frac{p}{1+p} m \Delta_{GDoF} + \sum_{j=1}^M m  \right) \log_Q \left ( \lfloor \frac{Q-1}{2}\rfloor -1 \right)  + o(m)
\end{align}
where (a) follows from $\sum_{j:\beta_j <2} h_j = \sum_{j:\beta_j>2}m(\beta_j-2)$. Now, we have the following bound on the GDoF;
\begin{align}
GDoF(\beta_1, \beta_2 \ldots \beta_M) & =  \limsup_{m\rightarrow \infty} \frac{C_{sym}(SNR,\beta_1,\ldots \beta_M)}{M\log_Q\left( Q^{2m}\right)} \nonumber\\
& \geq \lim_{m\rightarrow \infty} \frac{\left( \frac{p}{1+p} m \Delta_{GDoF} + \sum_{j=1}^M m  \right) \log_Q \left ( \lfloor \frac{Q-1}{2}\rfloor -1 \right)  + o(m)}{mM} \nonumber\\
&\stackrel{(a)} = \frac{ \frac{p}{1+p} \Delta_{GDoF} }{M} +1  
\end{align}
where (a) follows from large enough $Q$. The above inner bound matches outer bound (\ref{eq:GDoF_OB}) when $\Delta_{GDoF} < 0$ and this completes the GDoF characterization. 
\section*{Acknowledgment}
The work was supported in part by NSF awards 1136174 and 1314937.
Additionally, we gratefully acknowledge support by Intel and Verizon.

\appendix \label{sec:appendix}
\subsection{Proof of outer bound (\ref{eq:OB_4})} \label{sec:LD_OB_R1}
Using Fano's inequality for $Rx_1$, for any $\epsilon >0$, there exists a large enough $N$ such that;
\begin{align}
&NR^{(1)} - N\epsilon \nonumber\\ 
&\leq I(W^{(1)};\mathbf{Y}^{(1)}_{1:N}, \mathbf{S}_{1:N}) \nonumber\\
&= I(W^{(1)};\mathbf{Y}^{(1)}_{1:N}| \mathbf{S}_{1:N}) \nonumber\\
&\leq  H(\mathbf{Y}^{(1)}_{1:N}| \mathbf{S}_{1:N})  \nonumber\\
&\leq \sum_{t=1}^{N} H(\mathbf{Y}^{(1)}[t]| \mathbf{S}[t])  \nonumber\\
&=\sum_{t=1}^{N}  \sum_{\mathbf{s}} \mathbb{P}(\mathbf{S}[t]=\mathbf{s}) H(\mathbf{Y}^{(1)}[t]| \mathbf{S}[t]= \mathbf{s}) \nonumber\\
&\leq \sum_{t=1}^{N}  \sum_{\mathbf{s}} \mathbb{P}(\mathbf{S}[t]=\mathbf{s}) \sum_{j=1}^{M} 
n_j \mathbb{I}_{j \not \in \mathbf{s}} + \max(n_j,k_j) \mathbb{I}_{j \in \mathbf{s}} 
\nonumber\\
& = N\sum_{j=1}^{M} n_j + p (\max(n_j,k_j)-n_j)  \nonumber\\
&=N p\Delta +  N\sum_{j=1}^M n_j(1+p) - (n_j-k_j)^{+}p
\end{align}
where $\Delta =\displaystyle{\sum_{j=1}^{M}}  \max(n_j,k_j) + (n_j-k_j)^{+} - 2n_j $. The outer bound on $R^{(2)}$ follows by symmetry and this completes the proof of outer bound (\ref{eq:OB_4}).
\subsection{Proof of outer bound (\ref{eq:OB_3})} \label{sec:LD_OB_R1_R2_sum}
Using Fano's inequality for $Rx_1$ and $Rx_2$, for any $\epsilon >0$, there exists a large enough $N$ such that;
\begin{align}
&NR^{(1)} + N R^{(2)} - 2N\epsilon \nonumber\\
& \leq I(W^{(1)}; \mathbf{Y}^{(1)}_{1:N} , \mathbf{S}_{1:N}) +  I(W^{(2)}; W^{(1)}, \mathbf{Y}^{(1)}_{1:N}, \mathbf{Y}^{(2)}_{1:N}, \mathbf{S}_{1:N}) \nonumber\\ 
&= I(W^{(1)}; \mathbf{Y}^{(1)}_{1:N} | \mathbf{S}_{1:N}) + I(W^{(2)};  \mathbf{Y}^{(1)}_{1:N}, \mathbf{Y}^{(2)}_{1:N}  | \mathbf{S}_{1:N}, W^{(1)}) \nonumber \\
&=H(\mathbf{Y}^{(1)}_{1:N} | \mathbf{S}_{1:N}) - H(\mathbf{Y}^{(1)}_{1:N} | \mathbf{S}_{1:N}, W^{(1)})   +  H( \mathbf{Y}^{(1)}_{1:N}, \mathbf{Y}^{(2)}_{1:N}  | \mathbf{S}_{1:N} , W^{(1)})   \nonumber\\
&= H(\mathbf{Y}^{(1)}_{1:N} | \mathbf{S}_{1:N}) +  H(  \mathbf{Y}^{(2)}_{1:N} | \mathbf{Y}^{(1)}_{1:N},  \mathbf{S}_{1:N}, W^{(1)})  \nonumber\\
&= H(\mathbf{Y}^{(1)}_{1:N} | \mathbf{S}_{1:N})  +  H(   \hat{\mathbf{X}}^{(2)}_{1:N} | \mathbf{V}^{(1)}_{1:N},   \mathbf{S}_{1:N}, W^{(1)})  \nonumber\\
&\leq \sum_{t=1}^{N} H(\mathbf{Y}^{(1)}[t] | \mathbf{S}[t])  +  \sum_{t=1}^{N} H(  \hat{\mathbf{X}}^{(2)}[t] |\mathbf{V}^{(1)}_{\mathbf{S}[t]}[t] ,  \mathbf{S}[t] )  \nonumber \\
&= \sum_{t=1}^{N} \sum_{\mathbf{s}}  \mathbb{P}(\mathbf{S}[t]=\mathbf{s}) H(\mathbf{Y}^{(1)}[t] | \mathbf{S}[t]=\mathbf{s} )  + \sum_{t=1}^{N} \sum_{\mathbf{s}}  \mathbb{P}(\mathbf{S}[t]=\mathbf{s})  H(  \hat{\mathbf{X}}^{(2)}[t] | \mathbf{V}^{(1)}_{\mathbf{S}[t]}[t] ,  \mathbf{S}[t]=\mathbf{s})  \nonumber\\
&\leq  \sum_{t=1}^{N}  \sum_{\mathbf{s}}  \mathbb{P}(\mathbf{S}[t]=\mathbf{s})  \sum_{j=1}^{M} n_j \mathbb{I}_{j \not \in \mathbf{s} } + \max(n_j,k_j) \mathbb{I}_{j \in \mathbf{s} }   + \sum_{t=1}^{N} \sum_{\mathbf{s}}  \mathbb{P}(\mathbf{S}[t]=\mathbf{s})  \sum_{j=1}^{M} n_j \mathbb{I}_{j \not \in \mathbf{s}}   +  (n_j-k_j)^{+}  \mathbb{I}_{j \in \mathbf{s}}    \nonumber\\
&=  \sum_{t=1}^{N}    \sum_{j=1}^{M}   n_j  (1-p) + \max(n_j,k_j) p   + \sum_{t=1}^{N} \sum_{j=1}^{M}   n_j (1-p)    + (n_j-k_j)^{+} p  \nonumber \\
&=  Np\Delta  + 2N \sum_{j=1}^{M}  n_j 
\end{align}
where $\Delta =\displaystyle{\sum_{j=1}^{M}}  \max(n_j,k_j) + (n_j-k_j)^{+} - 2n_j $.
This completes the proof of outer bound~(\ref{eq:OB_3}).
\subsection{Proof of outer bound~(\ref{eq:OB_R^i_GN})} \label{sec:GN_R1_OB}
Using Fano's inequality for $Rx_1$, for any $\epsilon >0$, there exists a large enough $N$ such that;
\begin{align}
&NR^{(1)} - N\epsilon \nonumber\\ 
&\leq I(W^{(1)};\mathbf{Y}^{(1)}_{1:N}, \mathbf{Y}^{(2)}_{1:N}, W^{(2)},  \mathbf{S}_{1:N}) \nonumber\\
& = I(W^{(1)};\mathbf{Y}^{(1)}_{1:N}, \mathbf{Y}^{(2)}_{1:N} | W^{(2)}, \mathbf{S}_{1:N}) \nonumber\\
&=  h \left ( \mathbf{Y}^{(1)}_{1:N}, \mathbf{Y}^{(2)}_{1:N} | W^{(2)}, \mathbf{S}_{1:N} \right ) -  h \left ( \mathbf{Y}^{(1)}_{1:N}, \mathbf{Y}^{(2)}_{1:N} | W^{(1)}, W^{(2)} ,\mathbf{S}_{1:N} \right ) \nonumber\\
&=  h \left ( \mathbf{Y}^{(1)}_{1:N}, \mathbf{Y}^{(2)}_{1:N} | W^{(2)}, \mathbf{S}_{1:N} \right ) -  \sum_{t=1}^{N} h \left ( \mathbf{Y}^{(1)}[t] , \mathbf{Y}^{(2)} [t]  | \mathbf{Y}^{(2)}_{1:t-1}, \mathbf{Y}^{(1)}_{1:t-1} ,W^{(1)}, W^{(2)}, \mathbf{S}_{1:N} \right ) \nonumber\\
&=  h \left ( \mathbf{Y}^{(1)}_{1:N}, \mathbf{Y}^{(2)}_{1:N} | W^{(2)}, \mathbf{S}_{1:N} \right ) -  \sum_{t=1}^{N} h \left ( \mathbf{Z}^{(1)}[t] , \mathbf{Z}^{(2)} [t]  | \mathbf{Y}^{(2)}_{1:t-1}, \mathbf{Y}^{(1)}_{1:t-1} ,W^{(1)}, W^{(2)}, \mathbf{S}_{1:N} \right ) \nonumber\\
&=  h \left ( \mathbf{Y}^{(1)}_{1:N}, \mathbf{Y}^{(2)}_{1:N} | W^{(2)}, \mathbf{S}_{1:N} \right ) -  \sum_{t=1}^{N} h \left ( \mathbf{Z}^{(1)}[t] , \mathbf{Z}^{(2)} [t] \right ) \nonumber\\
&=  h \left ( \mathbf{Y}^{(1)}_{1:N}, \mathbf{Y}^{(2)}_{1:N} | W^{(2)}, \mathbf{S}_{1:N} \right ) - 2 N M \log \left ( \pi e   \right ) \nonumber\\
&=  \sum_{t=1}^{N} h \left ( \mathbf{Y}^{(1)}[t], \mathbf{Y}^{(2)}[t]  |  \mathbf{Y}^{(1)}_{1:t-1}, \mathbf{Y}^{(2)}_{1:t-1},  W^{(2)}, \mathbf{S}_{1:N} \right ) - 2 N M \log \left ( \pi e   \right ) \nonumber\\
& \leq   \sum_{t=1}^{N} h \left ( \mathbf{Y}^{(1)}[t], \mathbf{Y}^{(2)}[t]  |  \mathbf{Y}^{(1)}_{1:t-1}, \mathbf{Y}^{(2)}_{1:t-1},  W^{(2)}, \mathbf{S}[t] \right ) - 2 N M \log \left ( \pi e   \right ) \nonumber\\
&=  \sum_{t=1}^{N} \sum_{\mathbf{s}} \mathbb{P}(\mathbf{S}[t]=\mathbf{s}) h \left ( \mathbf{Y}^{(1)}[t], \mathbf{Y}^{(2)}[t]  |  \mathbf{Y}^{(1)}_{1:t-1}, \mathbf{Y}^{(2)}_{1:t-1},  W^{(2)}, \mathbf{S}[t]=\mathbf{s} \right ) - 2 N M \log \left ( \pi e   \right ) \nonumber\\
& \leq   \sum_{t=1}^{N} \sum_{\mathbf{s}} \mathbb{P}(\mathbf{S}[t]=\mathbf{s})  \sum_{j=1}^M  \mathbb{I}_{j \not \in \mathbf{s}} \left(  \log \left( \pi e \left(  1 + |g_{D,j}|^2 \right)  \right)  +  \log \left( \pi e \right)  \right) + \mathbb{I}_{j \in \mathbf{s}}  \log \left (  (\pi e)^2 \left( 1+ |g_{D,j}|^2 +|g_{I,j}|^2  \right ) \right )  - 2 N M \log \left ( \pi e   \right ) \nonumber\\
& = N \sum_{j=1}^M  (1-p) \log \left(  1 + |g_{D,j}|^2 \right)  + p \log \left( 1+ |g_{D,j}|^2 +|g_{I,j}|^2   \right ) 
\end{align} 
The bound on $R^{(2)}$ follows by symmetry and this completes the proof of outer bound (\ref{eq:OB_R^i_GN}). We also prove a looser bound on $R^{(i)}$ as shown below (the proof for this looser bound is used in the proof of outer bounds~(\ref{eq:causal_GN}) and (\ref{eq:OB_sum_GN})).
\par Using Fano's inequality for $Rx_1$, for any $\epsilon >0$, there exists a large enough $N$ such that;
\begin{align}
&NR^{(1)} - N\epsilon \nonumber\\ 
&\leq I(W^{(1)};\mathbf{Y}^{(1)}_{1:N}, \mathbf{S}_{1:N}) \nonumber\\
& = I(W^{(1)};\mathbf{Y}^{(1)}_{1:N}| \mathbf{S}_{1:N}) \nonumber\\
& =  h(\mathbf{Y}^{(1)}_{1:N}| \mathbf{S}_{1:N}) - h(\mathbf{Y}^{(1)}_{1:N}| W^{(1)}, \mathbf{S}_{1:N}) \nonumber\\
&\leq h(\mathbf{Y}^{(1)}_{1:N}| \mathbf{S}_{1:N}) - h(\mathbf{Y}^{(1)}_{1:N}| W^{(2)} ,W^{(1)}, \mathbf{S}_{1:N}) \nonumber\\
& = h(\mathbf{Y}^{(1)}_{1:N}| \mathbf{S}_{1:N}) -  \sum_{t=1}^{N} h(\mathbf{Y}^{(1)}[t]|\mathbf{Y}^{(1)}_{1:t-1} ,W^{(2)} ,W^{(1)} ,\mathbf{S}_{1:N}) \nonumber\\
&\leq h(\mathbf{Y}^{(1)}_{1:N}| \mathbf{S}_{1:N}) -  \sum_{t=1}^{N} h(\mathbf{Y}^{(1)}[t]|\mathbf{Y}^{(2)}_{1:t-1} , \mathbf{Y}^{(1)}_{1:t-1} ,W^{(2)} ,W^{(1)}, \mathbf{S}_{1:N}) \nonumber\\
&= h(\mathbf{Y}^{(1)}_{1:N}| \mathbf{S}_{1:N}) -  \sum_{t=1}^{N} h(\mathbf{Z}^{(1)}[t]|\mathbf{Y}^{(2)}_{1:t-1}  ,\mathbf{Y}^{(1)}_{1:t-1} ,W^{(2)} ,W^{(1)} ,\mathbf{S}_{1:N}) \nonumber\\
&= h(\mathbf{Y}^{(1)}_{1:N}| \mathbf{S}_{1:N}) -  \sum_{t=1}^{N} h(\mathbf{Z}^{(1)}[t]) \nonumber\\
&= h(\mathbf{Y}^{(1)}_{1:N}| \mathbf{S}_{1:N}) -  N M \log(\pi e) \nonumber\\
&\leq \sum_{t=1}^{N} h(\mathbf{Y}^{(1)}[t]| \mathbf{S}[t]) - N M \log(\pi e) \nonumber
\end{align}
\begin{align}
&=\sum_{t=1}^{N}  \sum_{\mathbf{s}} \mathbb{P}(\mathbf{S}[t]=\mathbf{s}) h(\mathbf{Y}^{(1)}[t]| \mathbf{S}[t]= \mathbf{s}) - N M \log(\pi e)  \nonumber\\ 
& \leq \left ( \sum_{t=1}^{N}  \sum_{\mathbf{s}} \mathbb{P}(\mathbf{S}[t]=\mathbf{s}) \sum_{j=1}^{M} 
\log(\pi e (1+|g_{D,j}|^2)) \mathbb{I}_{j \not \in \mathbf{s}} + \log \left ( \pi e  \left ( 1+ \left ( |g_{D,j}| + |g_{I,j}|  \right )^2  \right ) \right)  \mathbb{I}_{j \in \mathbf{s}} \right )- N M \log(\pi e) \nonumber\\ 
& = \left (\sum_{t=1}^{N}   \sum_{j=1}^{M} (1-p) \log \left ( \pi e \left ( 1+|g_{D,j}|^2 \right )  \right )  + p \log \left ( \pi e  \left ( 1+ \left ( |g_{D,j}| + |g_{I,j}|  \right )^2  \right ) \right)   \right )- N M \log(\pi e) \nonumber\\
& = N  \sum_{j=1}^{M} (1-p) \log \left ( 1+|g_{D,j}|^2   \right )  + p \log   \left ( 1+ \left ( |g_{D,j}| + |g_{I,j}|  \right )^2  \right) \label{eq:GN_looser_R1}
\end{align}
As mentioned above, this is a looser bound compared to (\ref{eq:OB_R^i_GN}), but the above proof is used in proving outer bounds (\ref{eq:causal_GN}) and (\ref{eq:OB_sum_GN}).
\subsection{Proof of outer bound (\ref{eq:OB_sum_GN})} \label{sec:OB_sum_GN}
Using Fano's inequality for $Rx_1$ and $Rx_2$, for any $\epsilon >0$, there exists a large enough $N$ such that;
\begin{align}
&NR^{(1)} + N R^{(2)} - 2N\epsilon \nonumber\\
& \leq I(W^{(1)}; \mathbf{Y}^{(1)}_{1:N} ,\mathbf{S}_{1:N}) +  I(W^{(2)}; W^{(1)}, \mathbf{Y}^{(1)}_{1:N}, \mathbf{Y}^{(2)}_{1:N}, \mathbf{S}_{1:N}) \nonumber\\ 
&= I(W^{(1)}; \mathbf{Y}^{(1)}_{1:N} | \mathbf{S}_{1:N}) + I(W^{(2)};  \mathbf{Y}^{(1)}_{1:N}, \mathbf{Y}^{(2)}_{1:N}  | \mathbf{S}_{1:N}, W^{(1)}) \nonumber\\
&=h(\mathbf{Y}^{(1)}_{1:N} | \mathbf{S}_{1:N}) - h(\mathbf{Y}^{(1)}_{1:N} | \mathbf{S}_{1:N} ,W^{(1)})   +  h( \mathbf{Y}^{(1)}_{1:N} ,\mathbf{Y}^{(2)}_{1:N}  | \mathbf{S}_{1:N}, W^{(1)}) - h( \mathbf{Y}^{(1)}_{1:N}, \mathbf{Y}^{(2)}_{1:N}  | \mathbf{S}_{1:N}, W^{(1)} ,W^{(2)})  \nonumber\\
&= h(\mathbf{Y}^{(1)}_{1:N} | \mathbf{S}_{1:N}) +  h(  \mathbf{Y}^{(2)}_{1:N} | \mathbf{Y}^{(1)}_{1:N} , \mathbf{S}_{1:N}, W^{(1)}) - h( \mathbf{Y}^{(1)}_{1:N}, \mathbf{Y}^{(2)}_{1:N}  | \mathbf{S}_{1:N} ,W^{(1)}, W^{(2)}) \nonumber\\
&= h(\mathbf{Y}^{(1)}_{1:N} | \mathbf{S}_{1:N}) +  h(  \mathbf{Y}^{(2)}_{1:N} | \mathbf{Y}^{(1)}_{1:N} , \mathbf{S}_{1:N}, W^{(1)}) - \sum_{t=1}^{N} h( \mathbf{Y}^{(1)}[t] ,\mathbf{Y}^{(2)}[t]  | \mathbf{Y}^{(1)}_{1:t-1}, \mathbf{Y}^{(2)}_{1:t-1} ,\mathbf{S}_{1:N}, W^{(1)} ,W^{(2)} ) \nonumber\\
&= h(\mathbf{Y}^{(1)}_{1:N} | \mathbf{S}_{1:N}) +  h(  \mathbf{Y}^{(2)}_{1:N} | \mathbf{Y}^{(1)}_{1:N} , \mathbf{S}_{1:N}, W^{(1)} ) - \sum_{t=1}^{N} h( \mathbf{Z}^{(1)}[t] ,\mathbf{Z}^{(2)} [t]  | \mathbf{Y}^{(1)}_{1:t-1}, \mathbf{Y}^{(2)}_{1:t-1},\mathbf{S}_{1:N}, W^{(1)}, W^{(2)}) \nonumber\\
&= h(\mathbf{Y}^{(1)}_{1:N} | \mathbf{S}_{1:N}) +  h(  \mathbf{Y}^{(2)}_{1:N} | \mathbf{Y}^{(1)}_{1:N} , \mathbf{S}_{1:N},W^{(1)}) - 2 N M \log(\pi e) \nonumber\\
&= h(\mathbf{Y}^{(1)}_{1:N} | \mathbf{S}_{1:N}) +  h( \hat{\mathbf{X}}^{(2)}_{1:N} \oplus \mathbf{Z}^{(2)}_{1:N} | \mathbf{Y}^{(1)}_{1:N}  ,\mathbf{S}_{1:N}, W^{(1)} ) - 2 N M \log(\pi e) \nonumber\\
&= h(\mathbf{Y}^{(1)}_{1:N} | \mathbf{S}_{1:N})  +  h( \hat{\mathbf{X}}^{(2)}_{1:N} \oplus \mathbf{Z}^{(2)}_{1:N}  | \mathbf{V}^{(1)}_{1:N}\oplus \mathbf{Z}^{(1)}_{1:N} , \mathbf{S}_{1:N} ,W^{(1)} )   - 2 N M \log(\pi e) \nonumber\\
&\stackrel{(a)} \leq N  \sum_{j=1}^{M} (1-p) \log \left ( 1+|g_{D,j}|^2   \right )  + p \log   \left ( 1+ \left ( |g_{D,j}| + |g_{I,j}|  \right )^2  \right)  \nonumber\\ & \quad +   h( \hat{\mathbf{X}}^{(2)}_{1:N} \oplus \mathbf{Z}^{(2)}_{1:N}  | \mathbf{V}^{(1)}_{1:N}\oplus \mathbf{Z}^{(1)}_{1:N} , \mathbf{S}_{1:N} ,W^{(1)} )  - N M \log(\pi e) \nonumber\\
& \leq N \sum_{j=1}^{M} (1-p) \log \left ( 1+|g_{D,j}|^2   \right )  + p \log   \left ( 1+ \left ( |g_{D,j}| + |g_{I,j}|  \right )^2  \right) \nonumber\\ & \quad +   \sum_{t=1}^{N} \sum_{\mathbf{s}} \mathbb{P}(\mathbf{S}[t]=\mathbf{s})  h( \hat{\mathbf{X}}^{(2)}[t] \oplus \mathbf{Z}^{(2)}[t]  | \mathbf{V}_{\mathbf{s}}^{(1)} [t]\oplus \mathbf{Z}^{(1)} [t] , W^{(1)} ,\mathbf{S}[t]=\mathbf{s} )  - N M \log(\pi e) \nonumber\\
& \leq N \sum_{j=1}^{M} (1-p) \log \left ( 1+|g_{D,j}|^2   \right )  + p \log   \left ( 1+ \left ( |g_{D,j}| + |g_{I,j}|  \right )^2  \right) \nonumber\\ & \quad +   \sum_{t=1}^{N} \sum_{\mathbf{s}} \mathbb{P}(\mathbf{S}[t]=\mathbf{s}) \sum_{j=1}^{M} \log(\pi e (1+|g_{D,j}|^2) )\mathbb{I}_{j \not \in \mathbf{s} } + \log \left ( \pi e \left ( 1+ \frac{|g_{D,j}|^2}{1+ |g_{I,j}|^2} \right ) \right ) \mathbb{I}_{j \in \mathbf{s}} - N M \log(\pi e) \nonumber\\
& = N \sum_{j=1}^{M} (1-p) \log \left ( 1+|g_{D,j}|^2   \right )  + p \log   \left ( 1+ \left ( |g_{D,j}| + |g_{I,j}|  \right )^2  \right) \nonumber\\ & \quad +  N \sum_{j=1}^{M}  (1-p)\log \left (1+|g_{D,j}|^2 \right ) + p \log \left ( 1+ \frac{|g_{D,j}|^2}{1+ |g_{I,j}|^2} \right )  \nonumber\\
& = N \left( 2 \sum_{j=1}^{M} \log \left ( 1+|g_{D,j}|^2   \right )  + p \Delta_G \right) 
\end{align}
where (a) follows from the proof of (\ref{eq:GN_looser_R1}) (see Appendix~\ref{sec:GN_R1_OB}) and $\Delta_G =  \sum_{j=1}^M  \log   \left ( 1+ \left ( |g_{D,j}| + |g_{I,j}|  \right )^2  \right)  +  \log \left (  1+ \frac{|g_{D,j}|^2}{1+ |g_{I,j}|^2} \right ) -2 \log \left ( 1+|g_{D,j}|^2   \right ) $.

\subsection{Achievability of corner points $D_1$ and $D_2$} \label{sec:D1D2_ach}
As shown in Figure~\ref{fig:crossover}, these corner points appear when $\Delta > 0$. We will describe the achievability of $D_1$ and achievability of $D_2$ follows by symmetry. The achievability of $D_1$ is similar to achieving $R_{NC}= \frac{p}{2}\Delta   + \sum_{j=1}^{M}  n_j$ (described in Section~\ref{sec:helping_mechanism}); with a slight modification for subcarriers with $\alpha_j >2$. The additive term $\frac{p}{2}\Delta $ appears in $R_{NC}$ because of bursty relaying in the leftover helper levels ($\Delta$ in number). For $D_1$, to achieve $R^{(1)}= p\Delta   + \sum_{j=1}^{M}  n_j$, we use an asymmetric version of bursty relaying as follows: In every block $Tx_1$ sends $N_B\Delta$ linear combinations of $pN_B\Delta$ fresh symbols in the leftover helper levels. $Rx_2$ receives $pN_B\Delta$ such linear combinations in every block; it recovers the constituent symbols and forwards them to $Tx_2$. In the next block, $Tx_2$ creates $N_B\Delta$ linear combinations of the constituent symbols sent by $Rx_1$ and sends them on its leftover helper levels. $Rx_1$ receives $pN_B\Delta$ of these linear combinations and thus recovers the constituent symbols. So compared to $R_{NC}$, $Rx_1$ now gains an additional rate $\frac{p}{2}\Delta$ but $Rx_2$ loses\footnote{The loss stems from $Tx_2$ not using its leftover helper levels for its own messages; it just uses them to relay messages for $Rx_1$.} rate $\frac{p}{2}\Delta$. This completes the achievability of $D_1$.
\subsection{Achievability of corner points $Q_1$ and $Q_2$} \label{sec:Q1Q2_ach}
Both $Q_1$ and $Q_2$ are achieved using a separation based scheme (\emph{i.e.,} no coding across subcarriers). We first describe the achievability of $Q_1$; achievability of $Q_2$ follows by symmetry. In $Q_1=(R^{(1)}, R^{(2)})= ( p \Delta  + \sum_{j=1}^{M} n_j(1+p) - (n_j-k_j)^+p ,\; \sum_{j=1}^M(n_j-k_j)^+ )$ we can rewrite rate $R^{(1)}$ as follows.
\begin{align}
&p \Delta  + \sum_{j=1}^{M} n_j(1+p) - (n_j-k_j)^+p \nonumber\\
&= \sum_{j:\alpha_j\leq 1} n_j + \sum_{j:\alpha_j > 1} n_j + (k_j-n_j)p \nonumber
\end{align}
Also, from the single carrier schemes in \cite{ihsiang_bursty_feedback}, the following rate tuples $(R^{(1)},R^{(2)})$ are achievable for a single carrier setup:
\begin{itemize}
\item $(n_j, n_j-k_j)$ for $\alpha_j \leq 1$.
\item $(n_j + (k_j-n_j)p , 0)$ for $\alpha_j > 1$.
\end{itemize}
Clearly, achieving the above rate tuple for each subcarrier and summing rates across subcarriers leads to corner point $Q_1$. The achievability of $Q_2$ follows by symmetry.
\end{document}